# Delayed phosphate release can highly improve energy efficiency of muscle contraction

Jiaxiang Xu[1#], Jiangke Tao[1#], Bin Chen[1,2*]

[1]Department of Engineering Mechanics, Zhejiang University, Hangzhou, China

[2]Key Laboratory of Soft Machines and Smart Devices of Zhejiang Province, Hangzhou, China

# These authors contributed equally.

* To whom correspondence should be addressed: chenb6@zju.edu.cn

**Abstract**

While the power stroke of myosin and the release of inorganic phosphate (Pi) play crucial roles in transforming ATP's chemical energy into mechanical work across diverse biological systems, the exact temporal relationship between these events continues to be intensely debated. In this study, from a functional perspective, we computationally investigate the impact of Pi-release kinetics during the power stroke on muscle contraction dynamics. By implementing a mechanics model of the sarcomere unit that comprehensively incorporates the chemomechanical cycle of individual myosin molecules, we successfully replicate a broad range of experimental observations through parameter variation. Our simulation results reveal that delayed Pi-release can significantly enhance energy efficiency during muscle contraction. This work suggests that a gradual Pi-release that is not directly coupled with the lever arm swing may offer a route to adjust the stability of a working myosin on the actin filament, thereby modulating the power stroke to influence muscle contraction.

## Introduction

The captivating dance of events within the chemomechanical cycle of a myosin during muscle contraction has been studied for many years and more of its mysteries are being revealed nowadays, as evidenced, for instance, by several recent works (1-6). Understanding the physiological sequences of these molecular events is not only crucial to elucidating the fundamental mechanisms that govern muscle dynamics, but also significant for treating relevant pathologies (7-9). Among these events, the power stroke and Pi-release can be key steps in the transduction from ATP chemical energy to the mechanical work (10-11). Early experiments employing photosensitive phosphate compounds (12-13) revealed a noticeable time interval before any reduction in force, which led to the hypothesis that the power stroke occurred immediately upon binding, followed by Pi-release (14). However, models representing Pi-release preceding the power stroke or the power stroke occurring prior to Pi-release were both capable of reproducing the reported relationship between filament load and velocity (15).

Employing a force spectroscopy technique with exceptionally high spatiotemporal resolution (16), researchers demonstrated that the power stroke rate is approximately one order of magnitude faster than Pi-release from myosin (17), indicating that the power stroke likely precedes Pi-release (3). Subsequent transient time-resolved FRET measurements further revealed that the rotation rate of myosin's lever arm exceeds that of Pi-release by about one order of magnitude (18). Meanwhile, Pi concentrations were found to have no significant impact on either the rate or magnitude of the power stroke in cardiac myosins (1), providing additional evidence for the precedence of the power stroke over Pi-release. Nevertheless, an alternative hypothesis suggests that Pi-release might still occur before the power stroke, with the released Pi remaining near ADP active sites rather than immediately entering the solution (5, 19), which would explain why Pi detection in the solution follows the power stroke. These conflicting results indicate that the precise temporal relationship between Pi-release and the power stroke continues to be a matter of substantial scientific debate (3).

In addition to detaching from the actin filament via the conventional ATP hydrolysis cycle, an attached myosin can also detach through unconventional forced bond breaking (16, 20-24). Based on detachment rates reported with the force spectroscopy technique (16), rates of bond breaking of a working myosin were extracted by considering coupling between bond breaking and state transition (25), which were found to vary strongly with nucleotide states and also forces. Among different nucleotide states, the bond formed at the AM.ADP.Pi state can be relatively weak (16). It was also reported that bonds formed between an attached myosin and actin filaments exhibit the counter-intuitive catch-bond behavior (23-24), characterized by an increase in lifetime in response to applied forces (26), which can provide a crucial means to stabilize attachment when needed (27-28).

In this investigation, from a functional perspective, we computationally analyze how Pi-release kinetics during the power stroke influence muscle contraction dynamics. To achieve this, we develop a mechanics model of the sarcomere unit incorporating the chemomechanical cycle of individual myosins. Through systematic simulations with varying parameters, we successfully reproduce a spectrum of muscle contraction characteristics that are consistent with experimental observations. Our simulation results reveal that the temporal decoupling between Pi-release and the lever arm swing within the power stroke, when synergistically with the stability of a working myosin, substantially enhances energy conversion efficiency. We furtherly suggest that this stability may be fundamentally linked to the mechanosensing pathways associated with Pi-release or ADP-release events. We believe that these findings provide novel insights into the regulatory mechanisms of Pi-release of myosin in muscle contraction, thereby advancing our comprehension of this complex biological process.

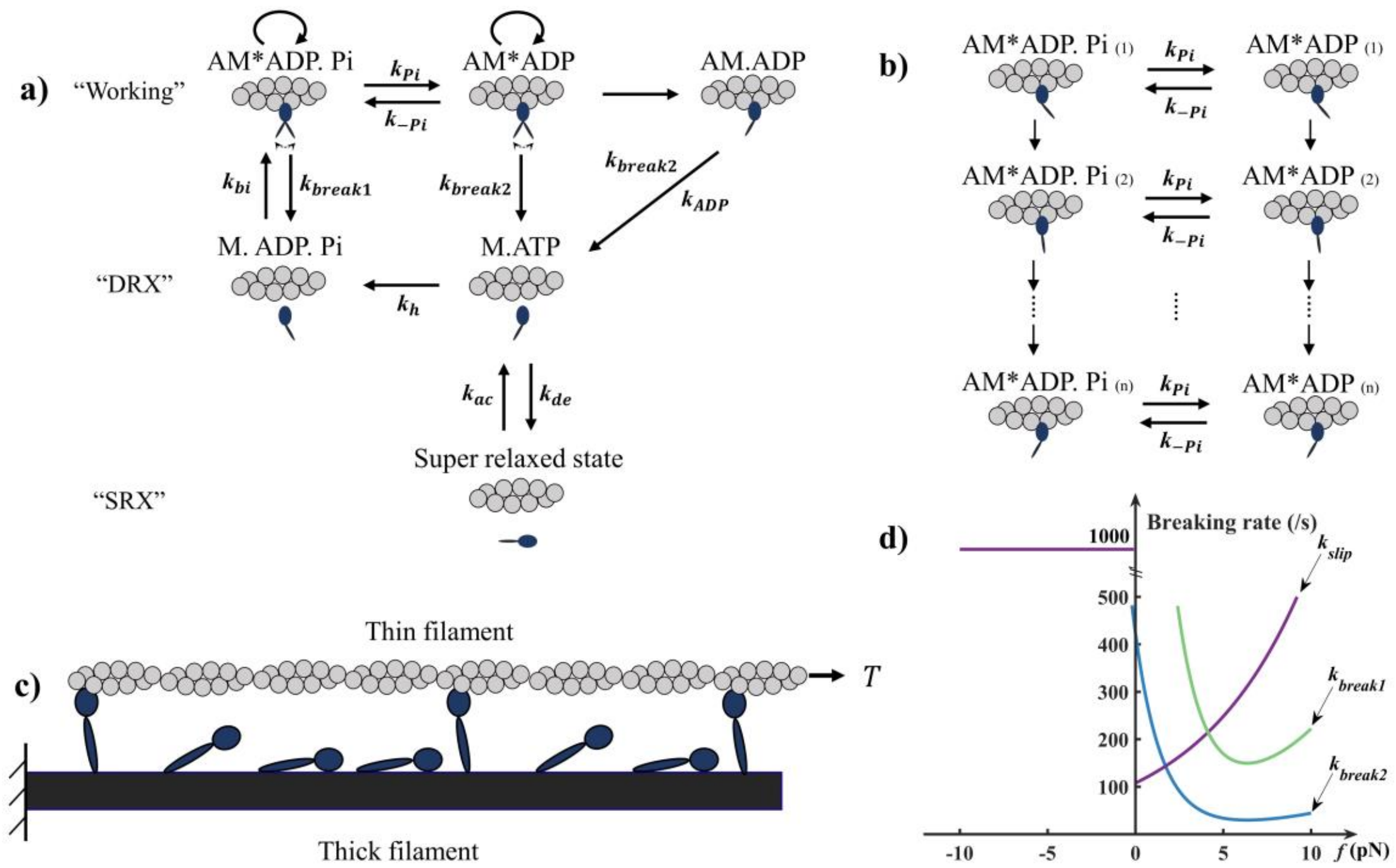

**Fig. 1** a) Six states are assigned and non-negligible transition rates among states are displayed. M.ATP and M.ADP.Pi are grouped as the DRX state; AM*ADP.Pi, AM*ADP, and AM.ADP are grouped as the Working state. At SRX, myosin is deactivated. Transition between SRX and M.ATP is regulated by filament load. Myosin moves from M.ATP to M.ADP.Pi upon hydrolysis and can then bind to the thin filament via Brownian motion. Binding leads to AM*ADP.Pi or AM*ADP with Pi-release that is not coupled to lever arm swing within the power stroke. At AM*ADP, Pi may rebind. At AM*ADP.Pi or AM*ADP, the lever arm can swing multiple times, as indicated by circular arrows back to each state, with rate and distance of each swing being regulated by motor force. When the accumulated swing distance due to one hydrolyzed ATP reaches the maximum, myosin releases ADP and captures an ATP to detach from actin filament. Alternatively, a motor can also detach directly from actin filament at any working state via bond breaking. If detaching at AM*ADP.Pi, it can rebind with ADP and Pi being still in its pocket. If detaching at AM*ADP or AM.ADP, it releases ADP and then catch an ATP to reach M.ATP state. b) The lever arm of myosin at AM*ADP.Pi or AM*ADP can swing multiple times among *n* possible sub-states within the power stroke. c) Illustration of a quarter of sarcomere unit for muscle contraction: the thin filament, the thick filament, and myosin motors, with a portion of motors being in SRX.

The thick filament is fixed on its right end and the thin filament is subjected to a filament load, $T$, on its left end. d) Bond breaking rates increase and then decrease with force for two catch-slip bonds, $k_{break1}$ and $k_{break2}$, adopted in the analysis by default. Bond breaking rate for a slip bond adopted in some analyses, $k_{slip}$, decreases with force for positive force and is set to be a large value of 1000/s for negative force.

**Model and numerical scheme**

a) Model

As illustrated in Fig. 1a, mainly based on the classical Lymn-Taylor theory (29), six states are assigned within the chemomechanical cycle of a single myosin, including SRX, M.ATP, M.ADP.Pi, AM*ADP.Pi, AM*ADP, and AM.ADP. Among them, M.ATP and M.ADP.Pi are grouped together as the DRX state and AM*ADP.Pi, AM*ADP, and AM.ADP are grouped together as the Working state, respectively.

At SRX, so called the super relaxed state (30), a myosin is deactivated and can't bind to the actin filament. The transition rate between SRX and M.ATP would be regulated by the filament load due to mechanosensing of the thick filament (31). A myosin transits from M.ATP to M.ADP.Pi when the hydrolyzation of ATP takes place. When a myosin is at M.ADP.Pi, it can bind to the thin filament through the Brownian motion.

When a myosin binds to the thin filament, the myosin will be at AM*ADP.Pi or AM*ADP with Pi-release. While it is controversial whether Pi-release occurs before or after the power stroke in the literature, we have assumed that Pi-release from the AM*ADP.Pi state is not directly coupled with the lever arm swing within the power stroke in the current work. At AM*ADP, Pi can rebind. In the model, at AM*ADP.Pi or at AM*ADP, the lever arm of a myosin can swing forward multiple times within the power stroke due to the transition among sub-states (32-36), as illustrated in Fig. 1b. Here asterisks emphasize that the lever arm can swing forward multiple times to continue releasing energy at these two states (32). Each swing can be temporarily arrested beyond an isomeric force, denoted as $f_a$. $f_a$ is set to be 6pN in our analysis (32, 34, 37), which can range from 1-10 pN (38-40) and may also be affected by temperature (41).

Note that, though multiple sub-states were assumed to exist within the power stroke (32-36), these sub-states have not been directly observed. Our model, rather than being limited to discrete sub-states, considers the smooth transition among numerous sub-states, which can be a good approximation when the swing distance associated with the conformational transition between real neighboring sub-states is relatively small. From this perspective, there exist more than six states illustrated in Fig. 1a in our model. Such a treatment in our model has substantially simplified the mathematical formulation of transitions among sub-states.

The rate and the distance of each swing at AM*ADP.Pi or at AM*ADP are regulated by force (32). When the motor force, denoted as $f$, $f < f_a$, the motor would swing forward at a rate given by $k_f = k_{f0}\exp\left(\frac{f_a - f}{f_a}\right)$ (32), where $k_{f0}$ is the forward swing rate at $f = f_a$; When $f > f_a$, the swing of the motor is assumed to be arrested. Note that the backward swing of the lever arm is neglected in the current model, which can be low for fast skeletal myosin but essential for cardiac myosin (4). The distance of each swing within the power stroke, denoted as $\Delta x$, is given by $\Delta x = \frac{f_a - f}{s_m}$, with $s_m$ being the spring constant of a motor.

When the accumulated swing distance of a myosin due to one hydrolyzed ATP reaches the maximum (32), denoted as $L_s$, the lever arm would not be able to swing any more and the motor will release the ADP and capture an ATP to quickly detach from the actin filament. Alternatively, a motor can detach from the actin filament at any working state through forced bond breaking. If the motor detaches through bond breaking at AM*ADP.Pi, it can rebind to the actin filament through the Brownian motion with ADP and Pi being still in its pocket (42). The total accumulated swing distance of the lever arm during multiple possible attachments due to one hydrolyzed ATP is also assumed to be less than $L_s$, beyond which the lever arm will not swing anymore. After a myosin detaches through forced bond breaking at AM*ADP or at AM.ADP, it is assumed to transit to the M.ATP state by quickly losing ADP and capturing an ATP.

Non-negligible transition rates among different states of a myosin considered in the model are displayed in Fig. 1a. The transition rate from SRX to M.ATP is denoted as $k_{ac}$. Since the exact thick filament activation mechanism is not fully understood, we tentatively employ the Bell's approximation (43) for this transition. With the Bell's formula (43), $k_{ac} = k_{01} \exp\left(\frac{T}{T_{01}}\right)$, where $k_{01}$ is the corresponding transition rate without filament load, denoted as $T$, and $T_{01}$ is a load scale. The backward transition rate from M.ATP to SRX is denoted as $k_{de}$. Also with the Bell's formula (43), $k_{de} = k_{10} \exp\left(-\frac{T}{T_{01}}\right)$, where $k_{10}$ is the corresponding transition rate without filament load.

The transition rate from M.ATP to M.ADP.Pi is denoted as $k_h$, which is set to be a constant. A working myosin is modeled as a passive linear spring in series with a rigid lever arm that can actively swing (32). The myosin binding rate from M.ADP.Pi to AM*ADP.Pi is given by $k_{bi} = \sqrt{\frac{\beta}{\pi}} 2\xi \exp(-\beta U^2) / [1 + \mathrm{erf}(0.3\sqrt{\beta})]$ (44), where $\beta = s_\mathrm{m} l_\mathrm{a}^2/(2k_\mathrm{B}T)$, with $s_m$ being the spring constant of the myosin, $k_B$ the Boltzmann constant, $T$ the absolute temperature, $l_a$ the spacing between neighboring binding sites on the thin filament, $U = u/l_\mathrm{a}$, with $u$ being the separation between the myosin and its binding site on the thin filament, and $\xi$ is a rate constant.

The Pi-release rate from AM*ADP.Pi denoted as $k_{Pi}$, is set to be a constant by default. The Pi-rebinding rate from AM*ADP, denoted as $k_{-Pi}$, is set to be zero by default. The relationship between motor force at AM*ADP.Pi or at AM*ADP, denoted as $f$, and motor stretch, denoted as $x$, is given by $f = s_m x$, with $x = u + d$, where $d$ is the accumulated swing distance of the lever arm, which is initially set to be zero upon the 1st attachment of a myosin with a hydrolyzed ATP to the thin filament.

The bond stability formed between a working myosin and actin varies with the nucleotide state (16, 24) and the bond formed at AM.ADP was suggested to be a catch-slip bond (24-25), with its lifetime counterintuitively increasing and finally decreasing with force. Importantly, the maximum of bond lifetime was found to be near the isomeric force of a myosin (24). In the current work, by default, we assume that the bond formed between myosin and actin at different working states is a catch-slip bond with the bond breaking rate, denoted as $k_{catch}$, being given by $k_{catch} = \vartheta\left(50\exp\left(-\frac{f}{1.5}\right) + \exp\left(\frac{f}{6}\right)\right)$ (44), where $f$ is of a unit of pN and $\vartheta$ is a rate constant. Note that the bond formed at AM*ADP.Pi might be relatively weak (1, 45). By default, as illustrated in Fig. 1d, the catch-slip bond breaking rate is set to be lower at AM*ADP or at AM.ADP, denoted as $k_{break2}$, than that at AM*ADP.Pi, denoted as $k_{break1}$, which is realized by assigning a smaller $\vartheta$ at AM*ADP or at AM.ADP, denoted as $\vartheta_2$, than that at AM*ADP.Pi, denoted as $\vartheta_1$. For comparison, a slip bond breaking rate used in some analyses is also displayed in Fig. 1d. The lumped kinetic rate from AM.ADP to M.ATP through the conventional ATP hydrolysis cycle is set to be a constant, denoted as $k_{ADP}$.

As illustrated in Fig. 1c, due to symmetry, only a quarter of a sarcomere unit is considered in developing the mechanics model for the muscle contraction (44). The model consists of the thin filament, the thick filament and multiple myosin motors uniformly distributed along the thick filament. The total number of myosins in the model is denoted as $N_m$ and the separation between neighboring myosins is denoted as $l_m$. Initially, a few "constitutively on" myosin motors, denoted as $N_a$, are at AM*ADP.Pi (31) and other myosin motors are set to be at SRX. Please note that "constitutively on" motors will never enter the SRX state, while other motors will randomly enter or leave SRX by following the kinetic scheme shown in Fig. 1a in our model. The thick filament is fixed on its left end while the thin filament is subjected to a filament load, *T*, on its right end. In the model, the thin filament in the sarcomere is modeled as an elastic rod with axial rigidity, $EA_{\text{thin}}$, and the thick filament is also modeled as an elastic rod with axial rigidity, $EA_{\text{thick}}$.

b) Numerical scheme

The numerical scheme employed in the simulations is a coupled Monte Carlo method and Finite-element method (44, 46). In the Finite-element method, the thin filament is discretized into one-dimensional two-node rod elements, each motor corresponds to a linear spring element, and the thick filament elasticity is also represented with one-dimensional two-node rod elements. A motor may change its state with time and the stiffness matrix of the whole structure and the nodal force vector are updated at each time step in the simulations. The displacement for each node is subsequently solved, and motor forces are obtained. The rates of all possible stochastic events are then calculated, based on which the time needed for the $i$th random event to take place, $\tau_i$, is obtained. The time needed for the next event to occur, $\Delta t$, would be the smallest among $\tau_i$. At the end of each time step, the system is updated, and the simulation proceeds to the next step (47). The program used for the analysis is self-coded and default values of parameters for the analysis are listed in Table 1. With chosen parameters in the analysis, simulations of our model can better represent skinned fibers from fast muscles. Model parameters in several analyses are adjusted to achieve qualitative consistency with experimental measurements.

**Table 1** Default values of parameters used in the simulations

| Parameter | Value | Parameter | Value |
|---|---|---|---|
| $N_m$ | 76 (48) | $N_a$ | 5 |
| $EA_{\text{thick}}$ | 46 nN | $EA_{\text{thin}}$ | 23 nN |
| $l_m$ | 14.3 nm (49) | $l_a$ | 5.5 nm (50) |
| $\vartheta_2$ | 8.3 /s (44) | $\vartheta_1$ | 41.5 /s |
| $\xi$ | 400 /s | $T_{01}$ | 40 pN |
| $k_{f0}$ | 1700 /s (48) | $s_m$ | 3 pN/nm (37) |
| $f_a$ | 6 pN (37) | $L_s$ | 6 nm (37) |
| $k_{01}$ | 200 /s | $k_{10}$ | 1000 /s |
| $k_{\text{ADP}}$ | 1000 /s (46) | $k_h$ | 1440 /s |
| $k_{Pi}$ | 250 /s (51) | $k_{-Pi}$ | 0 /s |
| $k_{\text{B}}T$ | $4.14\times10^{-21}$ J | $p_0$ | 83 pN. nm (52) |

In simulating the shortening distance at the loading end of the thin filament over time under given filament loads during steady-state muscle contraction, each simulation is terminated when the loading duration reaches 0.5 s or when the distance becomes excessively large. Linear fitting of the middle part of a simulated curve, where the shortening distance appears to vary linearly and stably with time, is performed to extract the shortening velocity. The extracted shortening velocities show low scatter and 10 runs are carried out to calculate the average of the shortening velocity at a given filament load. The output power is calculated by multiplying the shortening velocity with the given filament load. The power consumption is calculated by counting the total number of hydrolyzed ATPs per unit time multiplied by the chemical energy stored within one ATP. The energy efficiency is calculated by dividing the output power by the corresponding power consumption from ATP energy. The induced load per unit power consumption is calculated by dividing the filament load with the corresponding power consumption from ATP energy.

The average number of working motors in the shortening process is counted at a given filament load in the simulations. The force per working motor is calculated by dividing the filament load with the average number of working motors. Bond breaking occurrences/ATP is determined by dividing total number of bond breaking occurrences by total number of hydrolyzed ATPs in the shortening process at a given filament load. In the investigation of the [Pi] jump, the isometric loading condition at the initial [Pi] level is first simulated. Then, the [Pi] concentration is changed, and the simulation continues until the filament load stabilizes again. To calculate the average filament load over time resulting from the [Pi] jump, 100 simulation runs are performed.

**Results**

1) Effects of the Pi-release rate on muscle contraction

With the mechanics model described above, we are able to investigate the effect of the Pi-release rate on features of muscle contraction with results shown in Figs. 2-3. Note that the lower the Pi-release rate is, the more likely it is for the swing of the lever arm within a power stroke to proceed earlier than Pi-release. A special case is also considered in the simulation, where Pi-release instantly occurs upon the binding of a myosin to the actin filament, which can be regarded with an extremely high Pi-release rate. The predicted steady-state shortening velocities versus the filament loads are plotted in Fig. 2a. As seen from Fig. 2a, our model predicts that unloaded shortening velocities for $k_{Pi}$=50/s and $k_{Pi}$=250/s are very close, which are larger than that for $k_{Pi}$=1000/s or for the instant release. With the linear regression of simulated data close to zero shortening velocity, we determine the isometric load, which is 361pN for $k_{Pi}$=50/s, 383pN for $k_{Pi}$=250/s, 409pN for $k_{Pi}$=1000/s, and 415pN for instant release, respectively, which increases with the Pi-release rate. We have then employed a load value close to the predicted isomeric loads, denoted as $T_0$, 350pN, to scale all filament loads in Figs. 2-5. Please note that $T_0$ is not exactly the isometric load for $k_{Pi}$=50/s.

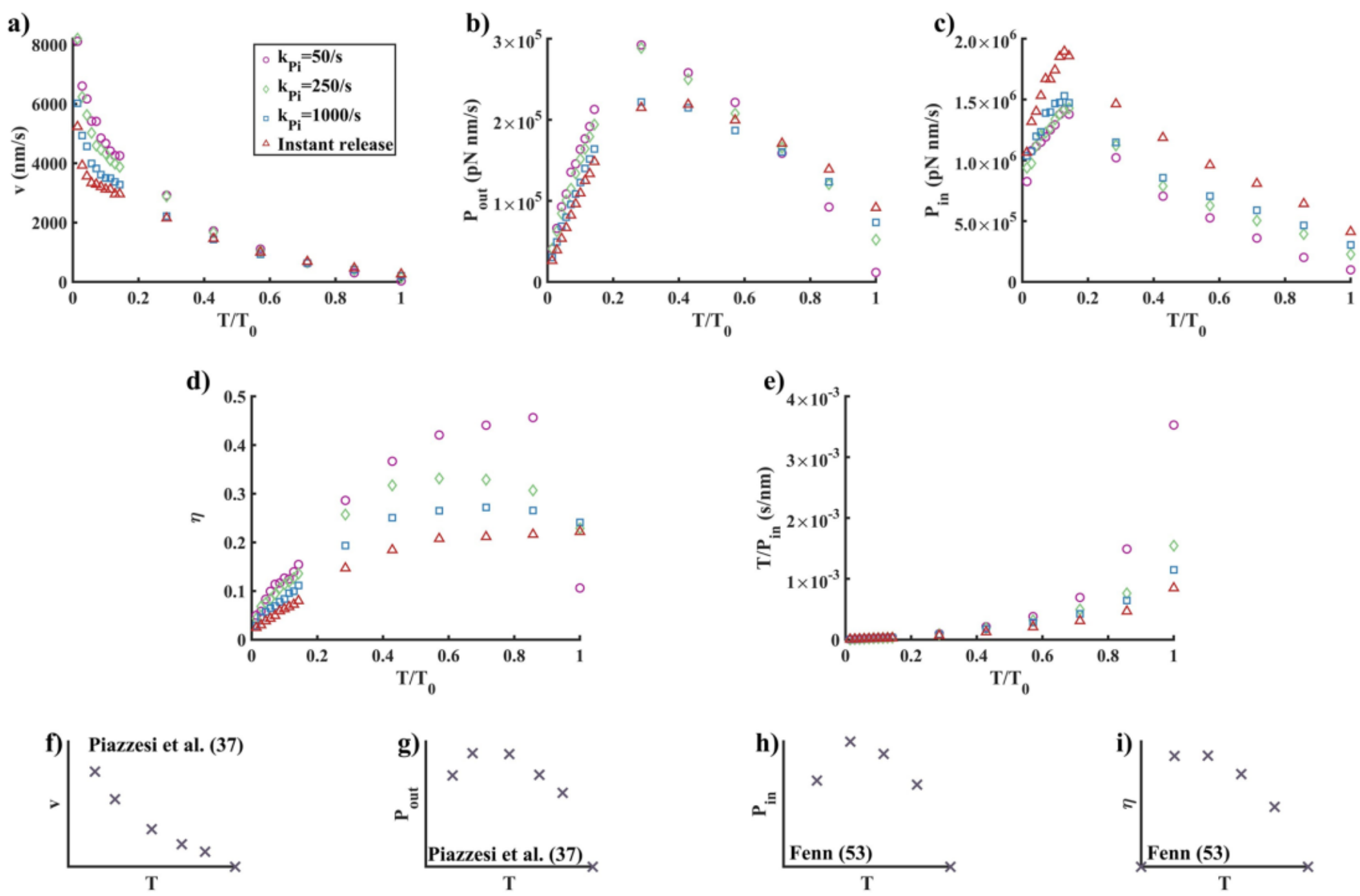

**Fig. 2** Predicted features of muscle contraction at different Pi-release rates (a-e): a) The shortening velocity versus the filament load; b) The output power versus the filament load; c) The power consumption versus the filament load; d) The energy efficiency versus the filament load; e) Induced load per unit power consumption versus the filament load. Reported features of muscle contraction from experiments (f-i): f) The shortening velocity versus the filament load (37); g) Calculated output power versus the filament load based on f); h) The power consumption versus the filament load (53); i) The energy efficiency versus the filament load (53).

In our simulations, a significant portion of working motors detach from the thin filament through bond breaking. In our model, with Pi being released, a working motor would transit from AM*ADP.Pi, which is set to be with a relatively higher bond breaking rate by default, to AM*ADP, which is set to be with a relatively lower bond breaking rate by default. If the Pi-release rate increases, this transition will occur earlier so that a working motor will be less likely to detach from the thin filament through the bond breaking, which can exert dragging force against the shortening of the thin filament, especially at low to moderate filament loads with relatively large shortening velocities. This would explain why the predicted shortening velocities filament loads are generally reduced as the Pi-release rate increases in Fig. 2a.

Based on the simulated shortening velocity versus the filament load, we furtherly calculate the output power during the steady-state muscle shortening, shown in Fig. 2b. As seen in Fig. 2b, the output power increases and then decreases with the filament load. Interestingly, its peak value occurs at a filament load being about 1/3 of the isometric load, in consistency with experiment (37, 54). We also calculate the power consumption during the steady-state shortening, denoted as $P_{in}$, considering the chemical energy stored within one ATP being $p_0$=83 pN.nm (52). As seen in Fig. 2c, the power consumption increases and then decreases with the filament load, in consistent with experiment (53). The calculated power consumption during muscle shortening is larger than that at the isometric condition, consistent with Fenn effect (53). Excitingly, when the Pi-release rate is higher than 250/s in our analysis, the power consumption corresponding to the maximal power output during the steady-state shortening is about 3 times of that at the isometric loading condition, in consistency with experiment (54). As seen From Figs. 2b,c, the Pi-release rate strongly affects both the output power and power consumption.

Based on results shown in Figs. 2b,c, we are able to calculate the energy efficiency, denoted as $\eta$, during muscle contraction. The energy efficiency during muscle contraction at different filament loads is shown in Fig. 2d. From Fig. 2d, it can be seen that the Pi-release rate strongly affects the energy efficiency, and increasing the Pi-release rate would substantially decrease the energy efficiency at low to moderate filament loads. This result is understandable. As the Pi-release rate gets lower, a myosin is more likely to detach through bond breaking with Pi being still in its pocket, which can then rebind to the actin filament and perform more power strokes to continue releasing the energy available from the hydrolysis of original ATP without consuming another ATP. As shown in Fig. 2d, the maximum efficiency predicted by our model is below 0.45 at different Pi release rates, which agrees well with reported values from the experiment, falling between 0.35 and 0.45 (55-57). Note that the energy efficiency exactly at the isometric load would be zero. In the close vicinity of the isometric load, the main function of the muscle contraction should sustain load instead to output power. Based on results shown on Fig. 2a, we then calculate the induced load per unit power consumption. As seen in Fig. 2e, it generally increases with the filament load.

We have also plotted reported data from experiments of the shortening velocity vs filament load (37) in Fig. 2f, the calculated output power versus the filament load based on data from Fig. 2f (37) in Fig. 2g, the power consumption versus the filament load (53) in Fig. 2h, and the energy efficiency versus the filament load (53) in Fig. 2i, which are in line with the trend of our simulation results shown in Figs. 2a-d, respectively.

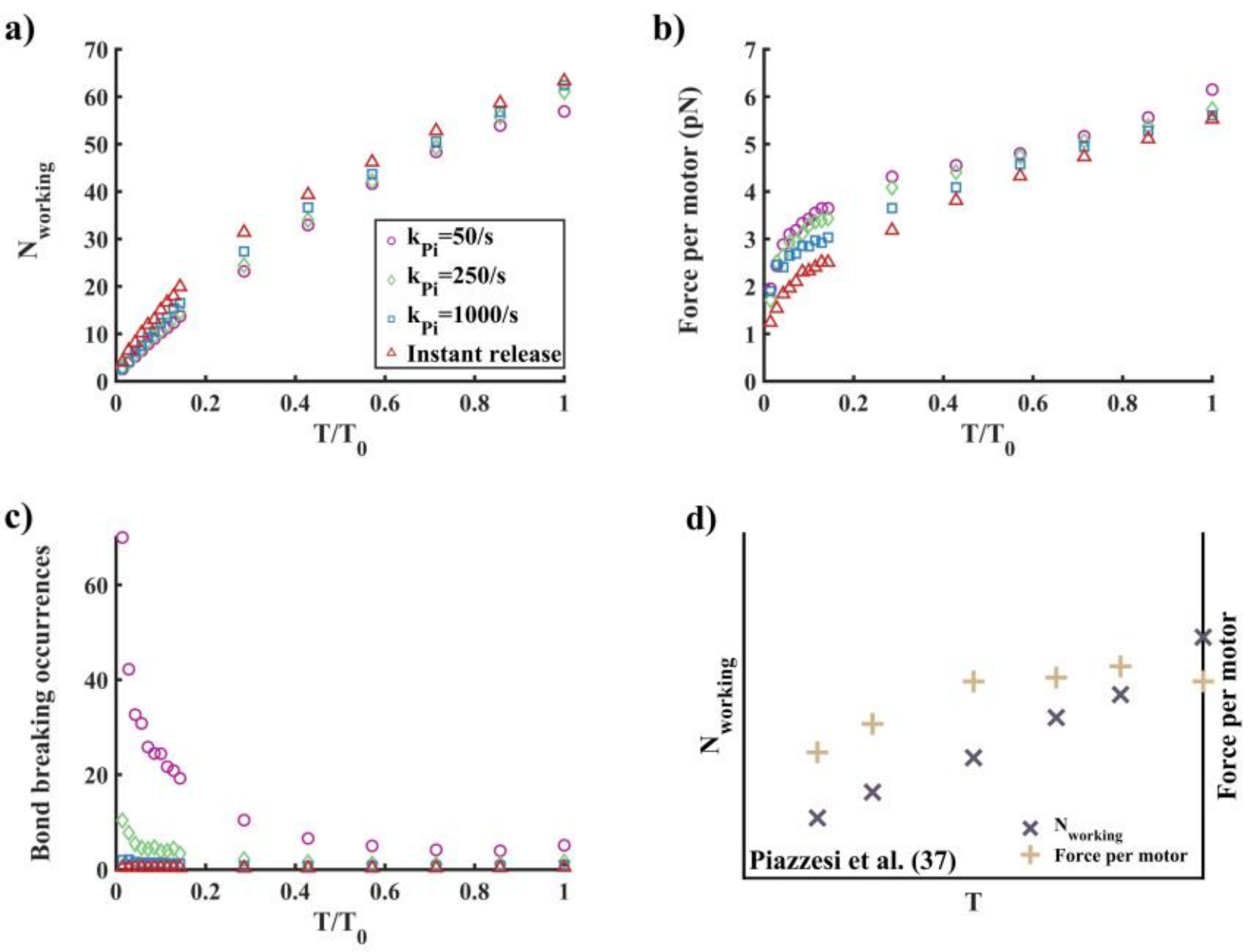

**Fig. 3** Predicted features of working motors at different Pi-release rates (a-c): a) Number of working motors versus the filament load; b) Force per working motor versus

the filament load; c) Bond breaking occurrences per ATP versus the filament load. d) Reported features of working motors from experiments (37): number of working motors versus the filament load and force per working motor versus the filament load.

Individual myosin motors were previously found to maintain a force of ~ 6 pN and the number of working motors was found to change in proportion to the filament load (37). We then predict the number of working motors versus the filament load, shown in Fig. 3a. In Fig. 3a, the number of working motors approximately linearly increases with the filament load for different Pi-release rates. Please note this result doesn't mean that the force-velocity relationship occurs due to changes in the number of working motors. The predicted force per working motor versus the filament load is shown in Fig. 3b, where the motor force at relatively high filament loads is regulated about 6 pN, which coincides with the isomeric force of a motor in our analysis. However, we would like to emphasize that the motor force at low filament loads is substantially lower than 6 pN in Fig. 3b, which indeed can also be seen from previous experiments (37). Small motor forces at low filament loads in our analysis are mainly due to that the relatively large shortening velocities at low filament loads would make the force on a working motor relatively small. Due to small motor forces at low filament loads, there can be multiple occurrences of bond breaking of a myosin with one hydrolyzed ATP, as seen in Fig. 3c, which increases as the Pi-release rate decreases. In Fig. 3d, we have also plotted number of working motors versus the filament load and force per working motor versus the filament load reported from experiments (37), which are in line with the trend of our simulation results shown in Figs. 3a-b.

Parametric analyses of $\xi$, $\vartheta_1$, and $k_{f0}$ have also been conducted and their effects on varied features of muscle contraction and working motors are shown in Figs. S1-S3, which suggest that our model exhibits robust qualitative behavior across a range of values for these parameters.

2) Effects of the stability of a working myosin on muscle contraction

Based on our predicted results shown in Fig. 3c, bond breaking of a working myosin may occur quite often in muscle contraction. To better understand the effect of bond breaking of a working myosin on muscle contraction, we decrease both catch-slip bond breaking rates to an extremely small value, denoted as "No breaking", or replace the catch-slip bond at AM*ADP.Pi with a slip bond while keeping the catch-slip bond at AM*ADP unchanged, denoted as "Mixed breaking", in the analysis. In our analysis, the breaking rate of a slip bond, denoted as $k_{slip}$, is given by the Bell's law (43), $k_{slip} = k_{slip}^0 e^{\frac{f}{f_b}}$, with $k_{slip}^0 = 107.9/\text{s}$ and $f_b = 6\text{pN}$, which monotonically increases with the bond force when $f > 0$ and is enforced to be $k_{slip}$=1000/s when $f <$ 0 to prevent dragging, as displayed in Fig. 1d.

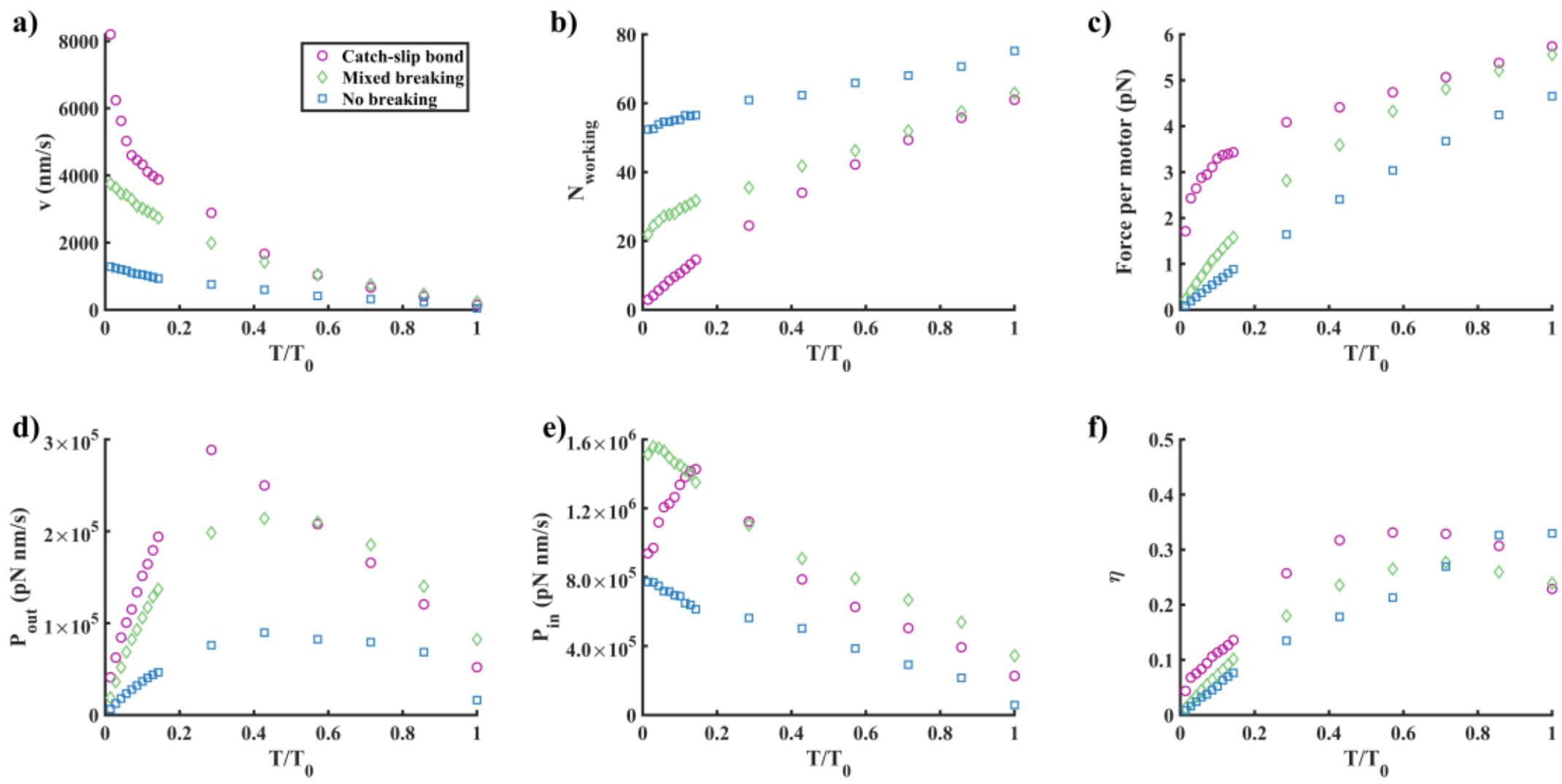

**Fig. 4** Predicted features of muscle contraction with "No breaking" or "Mixed breaking" compared to those with only catch-slip bond breaking: **a)** The shortening velocity versus the filament load; **b)** Number of working motors versus the filament load; **c)** Force per working motor versus the filament load; **d)** The output power versus the filament load; **e)** Power consumption versus the filament load; **f)** Energy efficiency versus the filament load.

Simulation results are shown in Fig. 4. Both "No breaking" and "Mixed breaking" substantially decrease the shortening velocity in Fig. 4a and also the output power of the muscle at low to moderate filament loads in Fig. 4d. The maximum of output power with "No breaking" or "Mixed breaking" is much lower than that with catch-slip bond breaking at both AM*ADP.Pi and AM*ADP and occurs about 50% of the isometric load, which is inconsistency with experiment (58). In Fig. 4b, the number of working motors for both "No breaking" and "Mixed breaking" only slightly varies with the filament load so that the motor force almost linearly increases with the filament load, which is inconsistency with experiment (37). Note that this result in the current analysis differs from the previous work, where a different model of a single myosin was employed (46). We also find that a significant portion of motors are subjected to a negative bond force at low filament loads for "No breaking". In Fig. 4e, the power consumption almost monotonically decreases with the filament load for both "No breaking" and "Mixed breaking", which is inconsistency with experiment (59). In Fig. 4f, the energy efficiency for "No breaking" or "Mixed breaking" at low to moderate filament loads is dramatically improved by catch-slip bonds at both AM*ADP.Pi and AM*ADP. These results suggest that particular bond stability is a crucial component in our model to qualitatively replicate experimental observations of muscle contraction.

3) Generalization capability of our mechanics model

Our mechanics model demonstrates strong generalization capability. With the model, we also simulate effects of both constant [Pi] and rapid jump in [Pi] on muscle contraction, as investigated in previous experiments. To do this, we furtherly assume that the effect of [Pi] is only to affect $k_{-Pi}$ in our model, given by the Michealis–Menten equation (60), $k_{-Pi} = k_{max}[\mathrm{Pi}]/(c_M + [\mathrm{Pi}])$, where $k_{max}$ is the maximal rebinding rate of Pi and $c_M$ is the Michaelis–Menten constant.

Our simulated effects of constant [Pi] on muscle contraction are displayed in Fig. 5. Adjusted parameters used in simulations are listed in Table 2. The catch-slip bond breaking rates in simulations are also adjusted, given by $k_{break1} = \begin{cases} 166/s\exp\left(-\frac{f}{3}\right) + 3.32/s\exp\left(\frac{f}{1.2}\right) & f \le 5pN \\ 178.45/\mathrm{s} + 166/s\exp\left(-\frac{f}{3}\right) + 12.45/s\exp(\frac{f}{5}) & f > 5pN \end{cases}$ and $k_{break2} = 199/s\exp\left(-\frac{f}{1.5}\right) + 8.3/s\exp\left(\frac{f}{5}\right)$, respectively, which are different from those provided by default. In Fig. 5, the horizontal axis, TB, represents the filament load normalized by the isometric load at 1 mM [Pi], which is calculated to be 275 pN. In Fig. 5a, the shortening velocities at different [Pi] are almost the same at very low filament loads, though the shortening velocity is generally lower at higher [Pi] at a given filament load. The isometric load is also lower at higher [Pi]. In Fig. 5b, the output powers at different [Pi] are almost same at very low filament loads, though the output power is generally lower at higher [Pi] at a given filament load. The maximum of the output power is much lower at higher [Pi]. In Fig. 5c, the motor force at low to medium filament loads and also its maximum appears to be almost the same at different [Pi]. In Fig. 5d, the energy efficiency is higher at higher [Pi] at low to medium filament loads. In Fig. 5e, the induced load per unit power consumption versus the filament load is generally higher at higher [Pi]. Reported data from previous experiments (33-34) are also plotted in Figs. 5a-c, showing that, simulation results based our model with varied parameters agree well with the experiment. Parametric analyses of $\xi$, $k_{f0}$, and $s_m$ have been conducted and their effects on varied features of muscle contraction at low or high constant [Pi] are shown in Figs. S4-S9, which suggest that our model exhibits robust qualitative behavior across a range of values for these parameters.

**Table 2** Values of parameters adjusted in the simulation of constant [Pi]

| Parameter | Value | Parameter | Value |
|---|---|---|---|
| $\xi$ | 200 /s | $k_h$ | 150 /s |
| $s_m$ | 2 pN/nm | $f_a$ | 5 pN |
| $c_M$ | 1.9 mM | $k_{\mathrm{ADP}}$ | 300 /s |
| $k_{max}$ | 1000 /s | | |

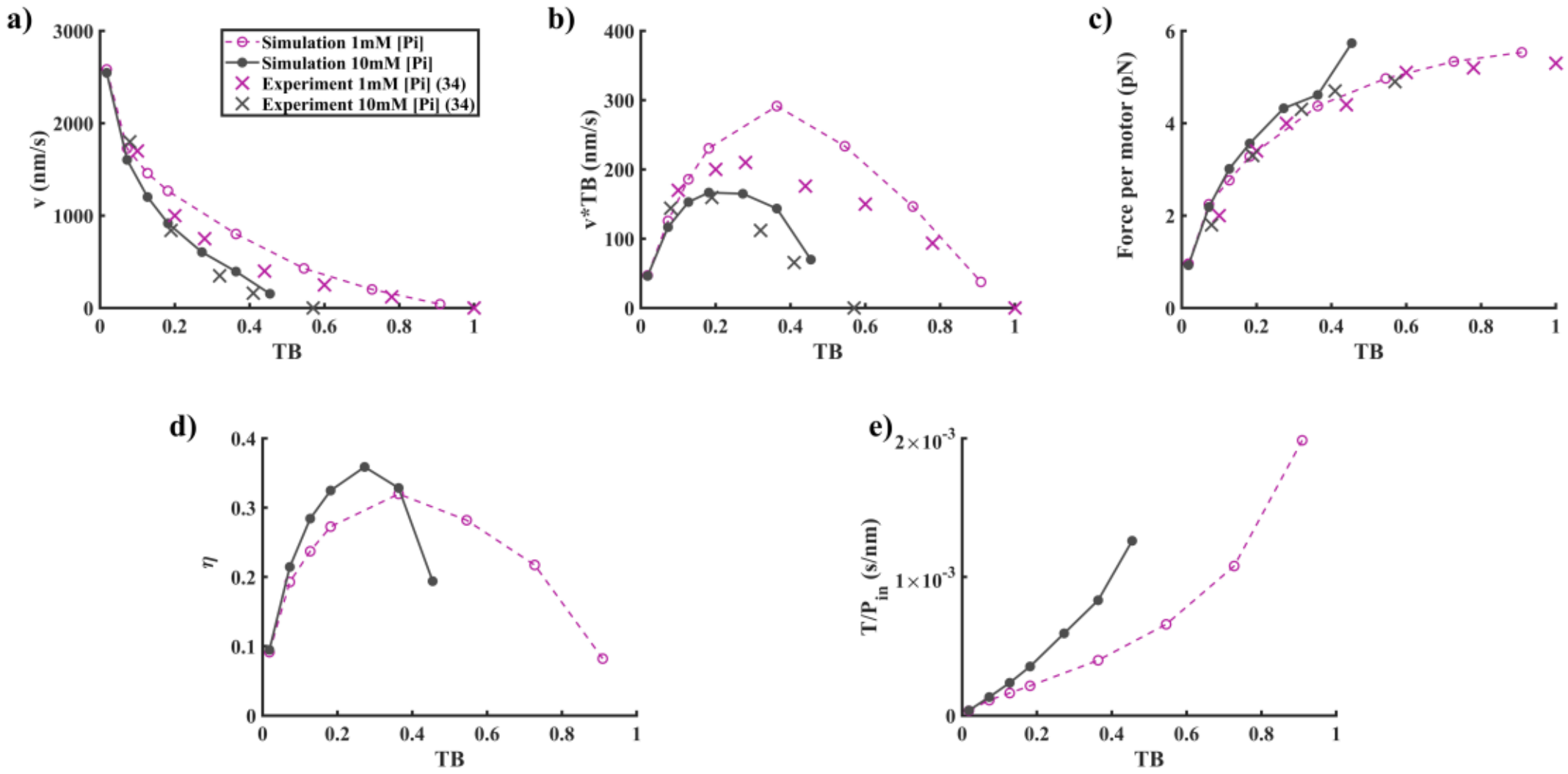

**Fig. 5** Simulated effects of constant [Pi] on features of muscle contraction: a) The shortening velocity versus the filament load; b) The output power versus the filament load; c) Force per working motor versus the filament load; d) The energy efficiency versus the filament load; e) Induced load per unit power consumption versus the filament load. Reported data from previous experiments (34) are plotted in (a-c).

We then simulate effects of rapid jump in [Pi] on muscle contraction with our model. As reported in the literature, the rate constant of tension decrease with a rapid increase in [Pi] was surprisingly different from that of tension increase with a rapid decrease in [Pi] within muscles (61-62). In our simulations, the rapid jump in [Pi] is superposed to the isometric loading condition. Adjusted parameters in simulations are listed in Table 3. Note that parameters utilized in the simulations for constant [Pi] and rapid jump in [Pi] are not exactly same. A transient release and then stretch at the same [Pi] of 2 mM superposed to the isometric loading condition is also simulated, with results shown in Fig. 6a. In Fig. 6b, an isometric load is generated at 20 mM [Pi], followed by a rapid decrease in [Pi] to 2 mM. In Fig. 6c, an isometric load is generated at 5 μMol of [Pi], followed by a rapid increase in [Pi] to 2 mM. With results shown in Fig. 6, we extract the rate constant of tension change due to the transient release and then stretch, denoted as $R_{TR}$, that following the rapid increase in [Pi], denoted as $R_{Pi(+)}$, and that following the rapid decrease in [Pi], denoted as $R_{Pi(-)}$, respectively. As listed in Table 4, the exacted rates are $R_{TR} =$ 1.81/s, $R_{Pi(+)}$= 3.18/s, and $R_{Pi(-)}$= 1.87/s, respectively, where $R_{Pi(-)}$ is close to $R_{TR}$ at same [Pi], while $R_{Pi(+)}$ is much higher than $R_{TR}$ at same [Pi]. As seen from Table 4, simulation results based on our model agree well with reported rates from previous experiments (61).

**Table 3** Values of parameters adjusted in the simulation of [Pi] jump

| Parameter | Value | Parameter | Value |
|---|---|---|---|
| $\vartheta_2$ | 0.005 /s | $\vartheta_1$ | 0.52 /s |
| $\xi$ | 1.25 /s | $k_{f0}$ | 53.13 /s |
| $k_{01}$ | 6.25 /s | $k_{10}$ | 31.25 /s |
| $k_{\mathrm{ADP}}$ | 3.13 /s | $k_h$ | 3.125 /s |
| $k_{Pi}$ | 7.8 /s | $k_{max}$ | 31.25 /s |
| $c_M$ | 1.9 mM | | |

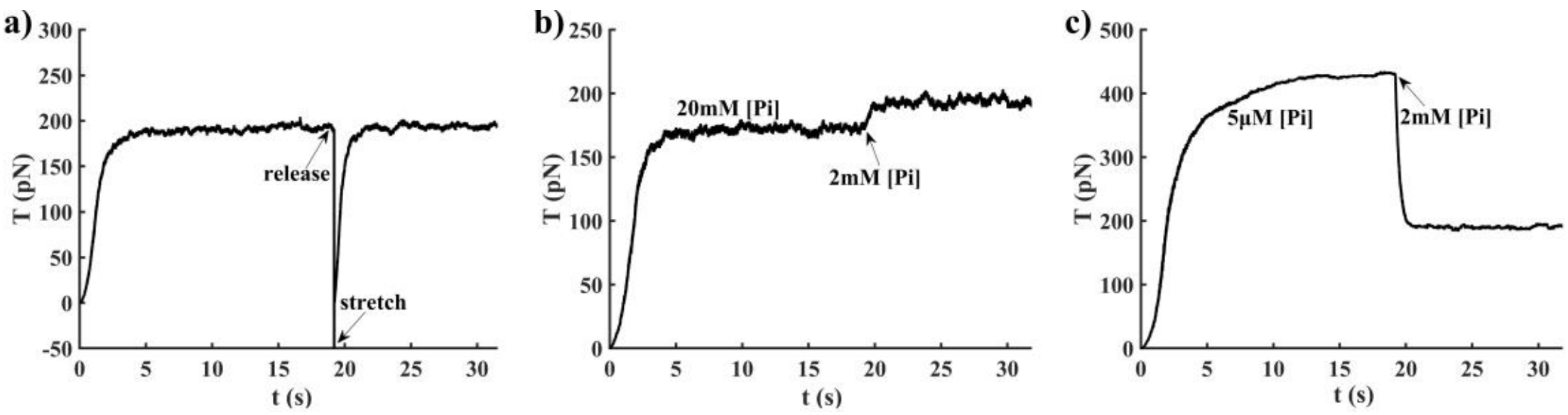

**Fig. 6** Predicted time course of the change in the isometric load due to a transient release and then stretch (a) or due to rapid jump in [Pi] (b, c): a) Load response of a sarcomere unit activated in 2 mM [Pi] solution and then subjected to a transient release and stretch; b) Load response of a sarcomere unit activated in 20 mM [Pi] solution and then subjected to a [Pi] jump to 2 mM; c) Load response of a sarcomere unit activated in 5 μM [Pi] and then subjected to a [Pi] jump to 2 mM.

**Table 4** Comparison of rate constants in [Pi] jump between simulation and experiment (62)

| | $R_{Pi(+)}(s^{-1})$<br>5μM→2mM | $R_{Pi(-)}(s^{-1})$<br>20mM→2mM | $R_{TR}(s^{-1})$<br>2mM |
|---|---|---|---|
| Simulation | 3.2 | 1.9 | 1.8 |
| Experiment (62) | 3.2±0.6 | 1.8±0.2 | 1.8±0.2 |

The asymmetric kinetic rates obtained in bi-directional rapid jumps of [Pi] in our simulations are closely related to varied bond breaking rates from different states of a working myosin. In our simulations, the bond breaking rate from the AM*ADP.Pi state is set to be much higher than that from AM.ADP state. At high [Pi] conditions, more motors would be at the AM*ADP.Pi state with a larger bond breaking rate, leading to a decrease in the total number of working motors and a subsequent reduction in the isometric load, as indicated in Fig. 6c. With a rapid increase in [Pi], a portion of myosins will transit directly from the AM*ADP state back to the AM*ADP.Pi state and then detach from the actin filament with bond breaking without going through the whole chemomechanical cycle, leading to a larger $R_{Pi(-)}$ than $R_{TR}$. Such a scenario appears to be consistent with the suggested sarcomere "give" in the literature (63). Interestingly, a biphasic force decay following an increase in [Pi] was observed in cardiac myofibrils (64). The transition from the slow first phase to the fast second phase was found to coincide with the onset of rapid lengthening in a single, weak sarcomere, which propagated from one sarcomere to the next along the myofibril in experiment (64). Such a propagation bears some similarities to the propagation of local structural instability related to material softening-then-hardening transition (65), which might have also been induced within sarcomeres due to a rapid increase in [Pi].

With a rapid decrease in [Pi], an extra portion of myosins would transit from the AM*ADP.Pi state to the AM*ADP state. Meanwhile, with a larger isometric load resulted from the decreased [Pi] at the steady state, an extra portion of myosins are needed to be recruited from either "DRX" state or "SRX" state, which would more or less go through the whole chemomechanical cycle of myosins. This would lead to a comparable $R_{Pi(+)}$ with $R_{TR}$ in our analysis. Such a scenario appears to be consistent with the previous suggestion that the force rise upon the decrease in [Pi] more closely reflects overall sarcomere cross-bridge kinetics (61).

We also simulate some other features of muscle contraction, which are compared well with those reported from experiments in the literature. For example, we simulate the variation of the number of active motors against the induced filament load under the isometric loading condition by letting only constitutively on motors be at the AM*ADP.Pi state and other motors be at SRX, initially. As seen in Fig. 7a, with $k_{ac}$ and $k_{de}$ both being regulated by the filament load, the induced filament load is observed to increase and more and more motors become active in our simulation, close to experiment (31).

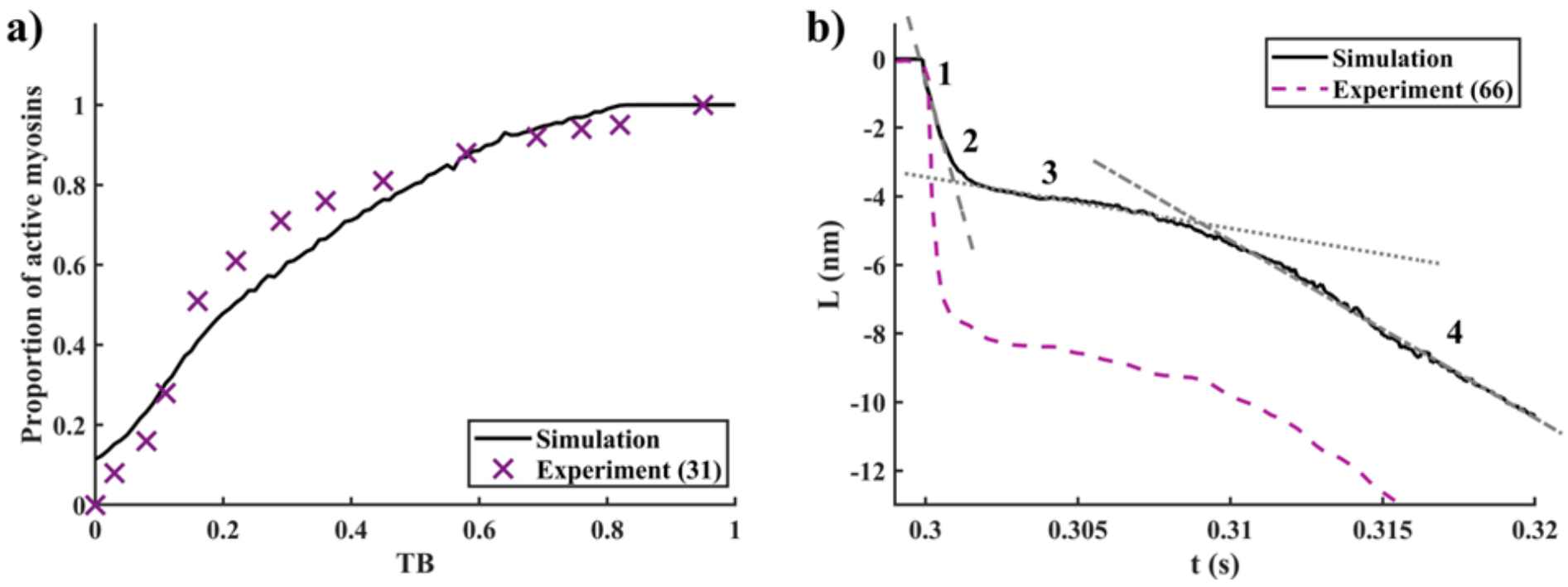

**Fig. 7** a) Proportion of active myosins varies with the induced filament load. Horizontal axis corresponds to the filament load normalized by the isometric value. The noise in the simulation results is due to the numerical error in Monte Carlo simulation with 100 runs. The adjusted parameters in the simulation are listed in Table 5. Experimental data (31) are plotted out for comparison. b) Simulated isotonic shortening distance following a stepwise drop of 50% in the filament load during isometric contraction displays with the typical four phases. The parameters adjusted in the simulations are listed in Table 2. Experimental data (66) are plotted out for comparison.

**Table 5** Values of parameters adjusted in the simulation of proportion of active myosins

| Parameter | Value | Parameter | Value | Parameter | Value |
|---|---|---|---|---|---|
| $k_{01}$ | 100 /s | $k_{10}$ | 2000 /s | $T_{01}$ | 30 pN |

In Fig. 7b, we simulate the isotonic shortening distance of a half sarcomere unit, denoted as $L$, following a stepwise 50% drop in the filament load during isometric contraction. As displayed in Fig. 7b, our simulation results agree with the previous experimental result (33) and exhibits with the typical four phases: an abrupt reduction of half-sarcomere length at phase 1 due to the half-sarcomere elasticity, a rapid shortening at phase 2 related to the synchronous execution of the power strokes of the working motors, a period of shortening at reduced velocity at phase 3 due to attachment and detachment of motors, and eventually shortening at a steady velocity at phase 4 also due to attachment and detachment of motors (66). Note that the difference at phase 1 between our simulation and experiment (33) can be due to that the passive elasticity from titin, Z-disc, M-line, etc. is not included in our model.

**Discussion**

1) Pertaining to catch bond mechanisms

As suggested from Fig. 4, the catch-slip bond breaking of a working myosin is an important feature in our model. As seen from Fig. 3b, the motor force is relatively small at low to moderate filament loads. With the catch-slip bond breaking adopted in our analysis, a working motor can quickly detach upon small motor forces so that it will not drag against a shortening filament at low to moderate filament loads. With Pi being still in the nucleotide pocket due to a relatively low Pi-release rate, the detached motor can then rebind to the actin filament to perform more power strokes. In this way, the energy efficiency in muscle contraction is improved at low to moderate filament loads. On the other hand, the motor force at high filament loads is regulated about the isometric force, which is close to the optimal value with the longest lifetime for the catch-slip bond breaking so as to sustain the filament load. With the synergy of Pi-release, the lifetime of the catch-slip bond breaking would be furtherly extended, as assumed in our model.

In our model, a working myosin detaching through the catch-slip bond breaking at AM*ADP or at AM.ADP is assumed to instantaneously transit to M.ATP state. In one way of thinking of this transition, a myosin detaches from the actin filament through forced bond breaking at AM*ADP or at AM.ADP (24), releases the ADP and then quickly catches an ATP. However, a problem would surface: the release rate of ADP from M.ADP is low, around 2/s (20), and M.ADP should not be a populated state within the cycle. This problem may be resolved if the release rate of ADP from M.ADP can increase with a portion of the energy being released.

Alternatively, it is known that the ADP-release rate at AM*ADP or at AM.ADP is force sensitive (67) and the ADP-release rate under the isometric condition can be several orders of magnitude lower than that without force (68-70). After the release of ADP, the motor will quickly secure an ATP and then detach from the actin filament. Such a transition might have been phenomenologically captured by the “catch” part of the catch-slip bond breaking in our model. Note that the transition in both ways would be expected to depend on ATP concentration in securing an ATP.

Though the catch-bond mechanism at AM*ADP seems to be relatively clear, that at AM*ADP.Pi is still elusive. In our model, the Pi-release rate from AM*ADP.Pi is assumed to be a constant. Consider that the AM*ADP state has higher stability than the AM*ADP.Pi state at relatively large motor forces on the thin filament. If the Pi-release rate from AM*ADP.Pi is force sensitive instead and increases with force, the promotion of the transition from the AM*ADP.Pi state to the more stable AM*ADP state by force can make the forced breaking from the AM*ADP.Pi state apparently behave as a catch-slip bond (71). Indeed, when we replace the catch-slip bond breaking at AM*ADP.Pi and constant Pi-release rate with slip bond breaking and force dependent Pi-release in the model, together with varied parameters, we find that our simulations can also agree well with experiment concerning effects of constant [Pi], as display in Fig. S10.

Interestingly, these discussions suggest that the detachment of a working motor from the actin filament as a catch-slip bond in our model can be physically related to the mechanosensing of Pi-release or ADP-release. On one hand, large motor forces promote the Pi-release to make detachment from the relatively weak state of AM*ADP.Pi apparently behave as a catch-slip bond. On the other hand, large motor force prevents the ADP-release from the relatively strong state of AM*ADP or AM.ADP to make detachment apparently behave as a catch bond. It seems that catch-slip bond breaking in our model might have provided a convenient way to capture the effects of mechanosensing of Pi-release or ADP-release within the power stroke on muscle contraction.

2) Comparison with previous works

Various models were developed to explain observed effects of Pi on muscle contraction. A five-state model of crossbridge kinetics with Pi-release occurring within the power stroke (45) was employed to reproduce experimental observations as [Pi] increases. However, those predictions (45) were based on the assumption that Pi's effect depends on the motor strain, which was not necessarily true (72). With the introduction of a branched pathway within the conventional crossbridge model (73), where a motor bound with Pi can detach, rapidly release hydrolysis products, and bind new ATP, the energetic cost of isometric contraction was successfully showed to increase with higher [Pi]. A slightly different branched kinetic model in which Pi induces detachment from a post-power stroke state was proposed, based on which effects of Pi on the force-producing capability of a small group of myosin molecules, as well as on the isometric ATPase in muscle fibers were explained (74).

Interestingly, models without branched kinetic pathways (69, 75) were also able to explain the relationships between [Pi], isometric force, and shortening velocity at low loads, though these models were found to produce unusual force-velocity relationships at high [Pi] and loads (76). With an ultra-high-speed optical trap to directly observe the timing and magnitude of the working stroke in β-cardiac myosin, it was showed that the initial actomyosin state can withstand high loads and detach before releasing Pi through a branched pathway (1). A model was then proposed, which involved short-lived, weakly-bound attachments that directly precede the working stroke, followed by Pi release and potential rebinding (1).

To elucidate the high power output observed during muscle shortening, it was proposed (33-34) that a working myosin, whether bound to both Pi and ADP or solely ADP, can slip from the original actin monomer to the next one, together with branched kinetic pathways. This model (33-34) introduced multiple sub-states within the power stroke (35-36) and treated mechanical and chemical events as orthogonal processes so that Pi release can occur at any stage of the power stroke. By incorporating these principles, together with the assumption that the release rates of Pi and ADP were conformation-dependent or force-dependent, the model explained isotonic velocity transients, reproduced the steady-state relationship between filament load, shortening velocity, and power (33), and provided a mechanistic explanation for the observed correlation between the rate of energy liberation and filament load during muscle contraction under constant [Pi] conditions (34).

In comparison, our model also incorporates multiple sub-states within the power stroke (35-36) and allows Pi release to occur at any of these states. However, quite a few distinctive features differentiate our current model from previous approaches (33-34). Our model permits a detached myosin from AM*ADP.Pi to rebind multiple times, enabling additional lever arm swing to release energy without requiring new ATP binding. Our model implements a continuous transition mechanism for lever arm swing during the power stroke (32), regulated by the difference between motor force and the isometric force (32). Our model does not require conformation-dependent Pi or ADP release, instead suggesting a potential force-dependent mechanosensing mechanism, as inferred from our discussed catch bond mechanisms. Our model is simulated through coupled Monte Carlo and Finite Element methods with robust extensibility. Through parameter variation, our simulations successfully reproduce both the effects of constant [Pi] and rapid [Pi] jumps, along with other characteristic features of muscle contraction – a strong generalization capability not previously achieved in published studies, to the best of our knowledge. Crucially, our relatively parsimonious model suggests that gradual Pi-release that is not directly coupled with the lever arm swing may provide a mechanism for tuning the stability of a working myosin on actin filaments, thereby modulating power stroke dynamics to influence muscle performance.

3) Shortcomings and limitations

While our model and analyses demonstrate notable capabilities, here we also acknowledge shortcomings and limitations to provide better understanding of model applicability and potential areas for future enhancement.

(i) To align the model with experimental observations under specific conditions, we have adjusted a range of parameters in the analysis. While this approach provides a reasonable fit, it may introduce uncertainties in the generalizability of the model to other experimental conditions.

(ii) The passive elasticity arising from Titin, M-line, Z-disk, etc., is omitted in our model. This omission contributes to discrepancies between the predicted and observed abrupt reduction in half-sarcomere length at phase 1 in the isotonic shortening following a stepwise drop in the filament load during isometric contraction.

(iii) We have neglected the backward rate of ATP hydrolysis off actin in the model for simplification, which was estimated to be several times lower than the forward rate (78). This simplification may limit the model's ability to capture certain physiological nuances.

(iv) The way that we have specified reaction rates to and from the SRX state in the model is not consistent with ATPase experiments (78). There might exist highly cooperative thick filament activation (78), which has not been considered in our model and may limit the model's ability to fully capture thick filament regulation mechanism.

(v) The degree to which the partial release of the energy from a detached myosin modulates the reverse rate of ATP hydrolysis and the dissociation kinetics of hydrolysis products (ADP and Pi) remains an area of uncertainty, which can affect energy efficiency, ATPase activity at isometric load, etc.

(vi) The rates of multiple possible rebinding of a detached Pi-bound myosin are assumed to be the same in our model. When considering that this rebinding rate is affected in simulations, for example, by the accumulated swing distance of the power stroke, both the shortening velocity and the energy efficiency will get lower, while the impact on the motor force or the power consumption is relatively small, as displayed in

Fig. S11. Unfortunately, the factors influencing the rebinding rate of a detached Pi-bound myosin from AM*ADP.Pi are unclear. This aspect warrants further investigation to refine the model's predictive accuracy.

(vii) The mechanism underlying catch bond breaking at AM*ADP.Pi remains unresolved. Specifically, whether this phenomenon is driven by mechanosensing of Pi-release requires experimental validation.

(viii) The stability of a Pi-bound working myosin on the actin filament is assumed to be only affected by force in the model. However, it remains unclear whether a multi-step Pi dissociation process (5) contributes to fine-tuning the stability of a Pi-bound working myosin, and also modulating the rebinding kinetics of a detached Pi-bound myosin.

**Conclusion**

With a mechanics model of a sarcomere unit, we simulate various features of muscle contraction in silico and results agree well with experiment. Our analysis reveals that relatively slow Pi-release that is not directly coupled with the lever arm swing can lead to highly improved energy efficiency in muscle contraction. We suggest that a gradual Pi-release may help tune the stability between a working motor and actin to modulate the power stroke so as to regulate overall performance of muscle contraction. This stability may be fundamentally linked to the mechanosensing of Pi-release or ADP-release. We believe that this work offers significant insights into the kinetics and functions of molecular-scale events governing force generation and motion regulation in multi-myosin systems, with potential applications extending to other classes of molecular motors.

**Acknowledgements**

This work was supported by the National Natural Science Foundation of China (Grant No.: 12372318) and Zhejiang Provincial Natural Science Foundation of China (Grant No.: LZ23A020004).

## Author contributions

B.C. designed research, J.X., J.T., and B.C. performed research, J.X., J.T., and B.C. analyzed data, J.X., J.T., and B.C. wrote the manuscript.

## Declaration of interests

The authors declare no competing interests and that Mr. Jiangke Tao is now a PhD candidate at the University of Macau.

## Supplementary Information

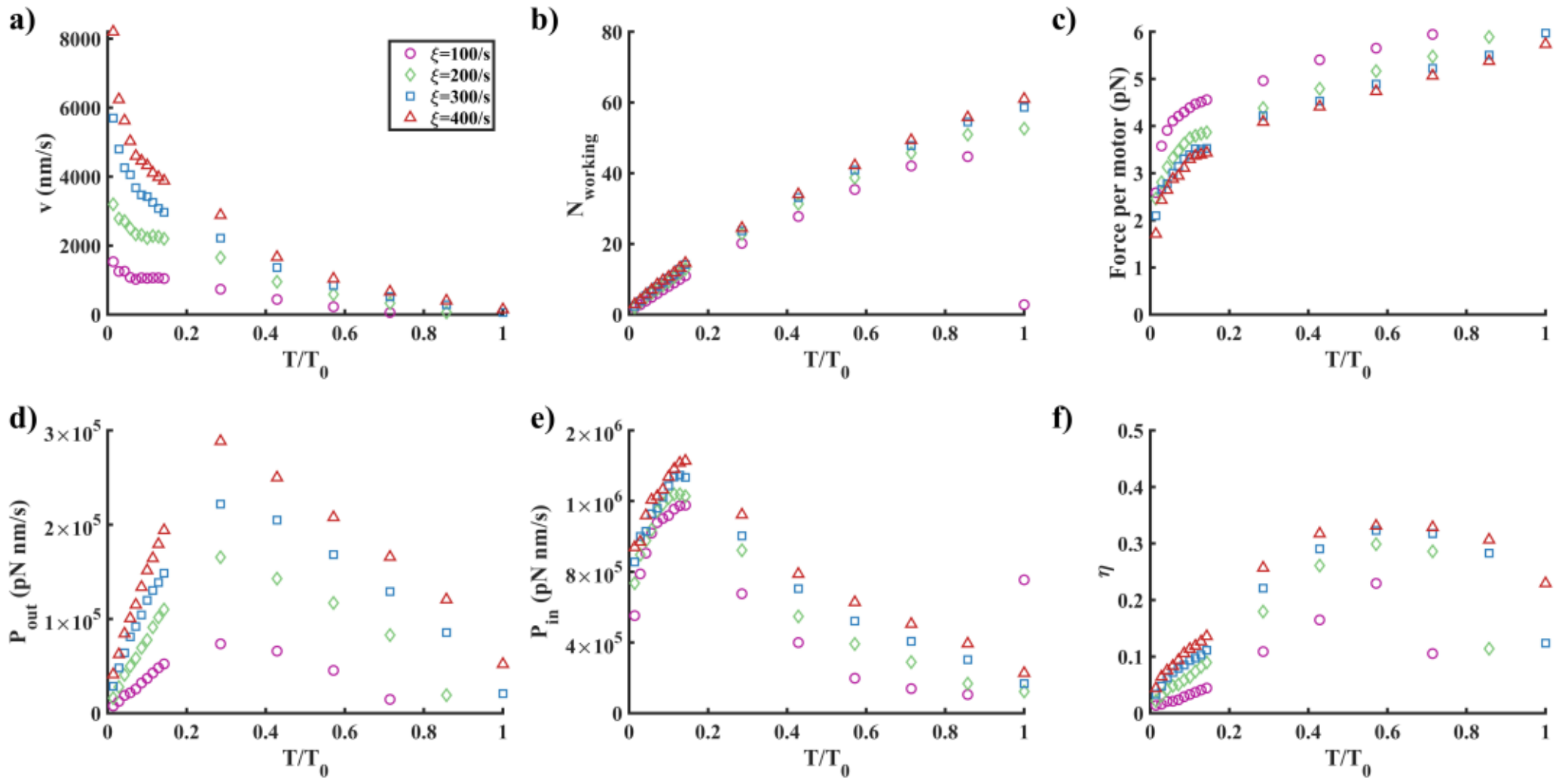

**Fig. S1** Predicted effects of $\xi$ on features of muscle contraction: **a)** The shortening velocity versus the filament load; **b)** Number of working motors versus the filament load; **c)** Force per working motor versus the filament load; **d)** The output power versus the filament load; **e)** Power consumption versus the filament load; **f)** Energy efficiency versus the filament load.

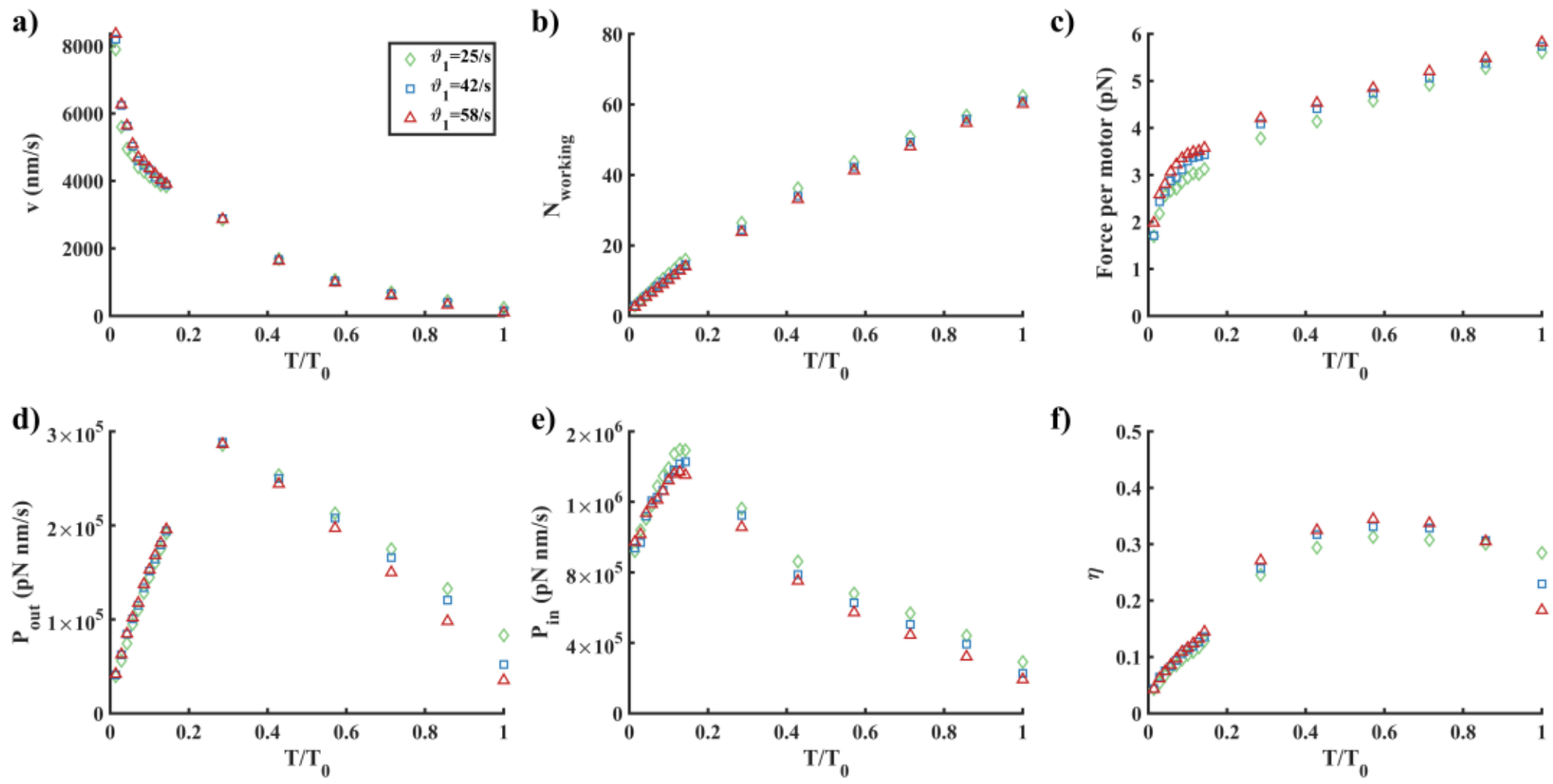

**Fig. S2** Predicted effects of $\vartheta_1$ on features of muscle contraction: **a)** The shortening velocity versus the filament load; **b)** Number of working motors versus the filament load; **c)** Force per working motor versus the filament load; **d)** The output power versus the filament load; **e)** Power consumption versus the filament load; **f)** Energy efficiency versus the filament load.

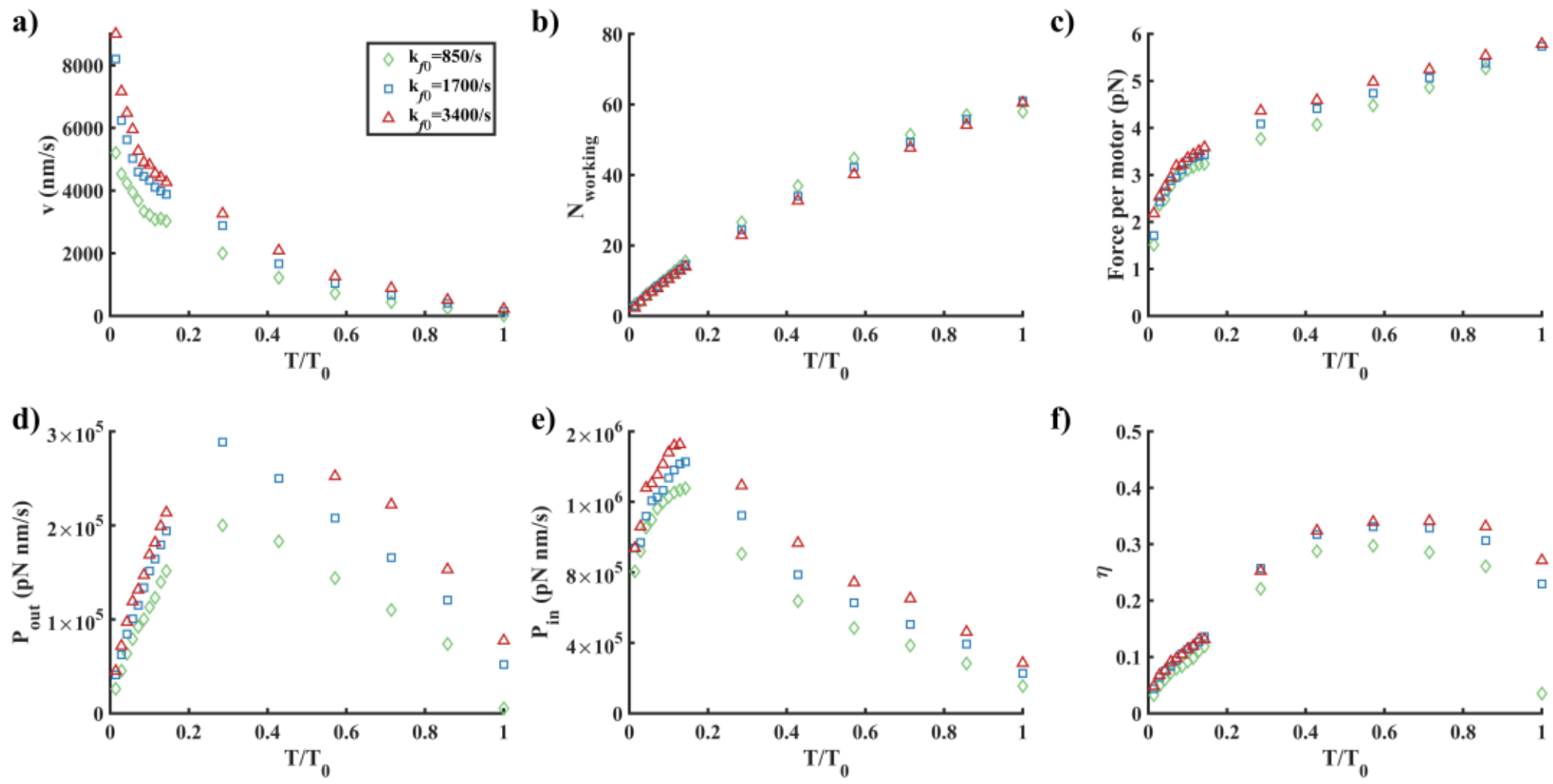

**Fig. S3** Predicted effects of $k_{f0}$ on features of muscle contraction: **a)** The shortening velocity versus the filament load; **b)** Number of working motors versus the filament load; **c)** Force per working motor versus the filament load; **d)** The output power versus the filament load; **e)** Power consumption versus the filament load; **f)** Energy efficiency versus the filament load.

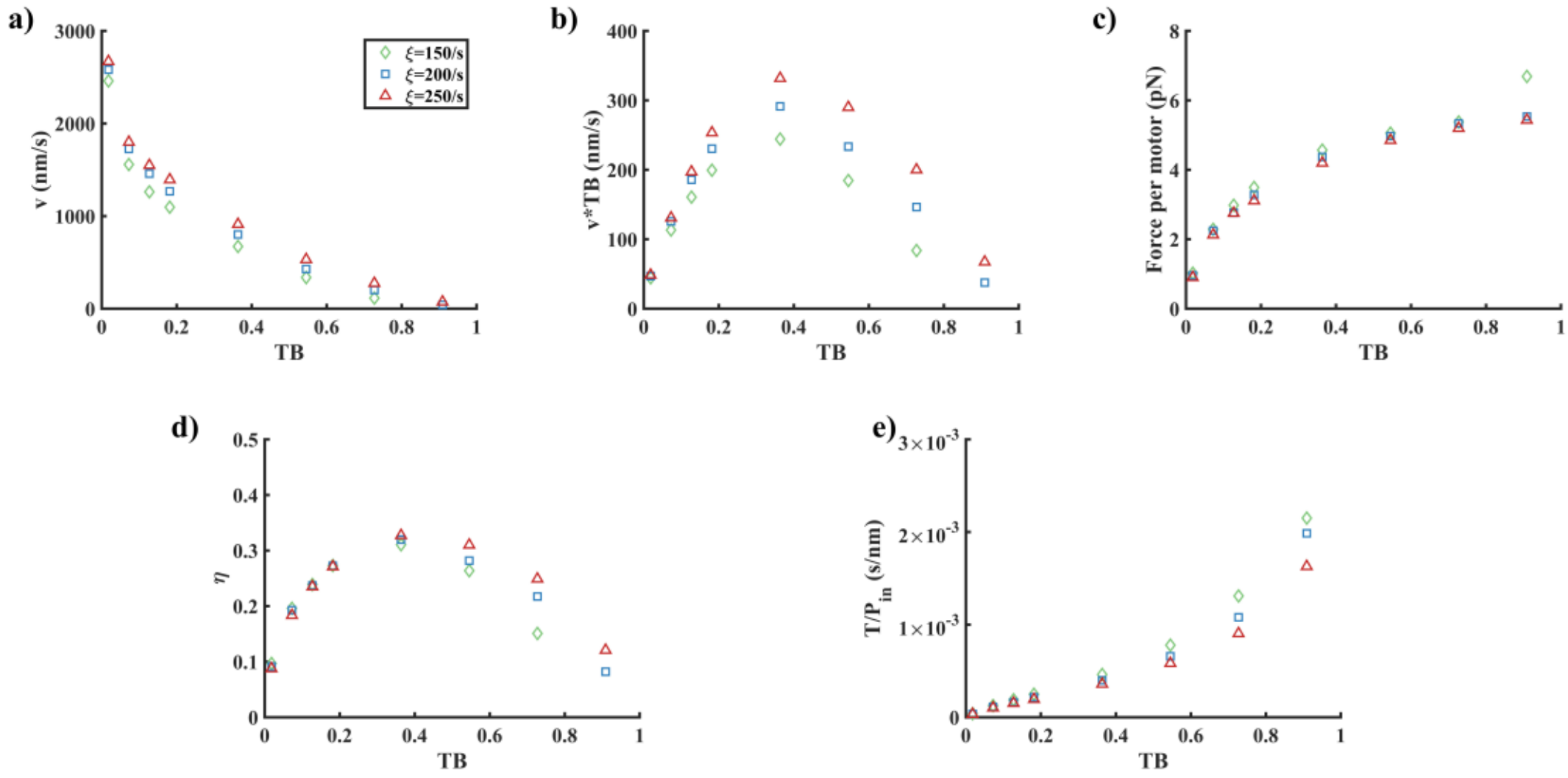

**Fig. S4** Predicted effects of $\xi$ on features of muscle contraction at constant [Pi] ([Pi] = 1mM): **a)** The shortening velocity versus the filament load; **b)** The output power versus the filament load; **c)** Force per working motor versus the filament load; **d)** The energy efficiency versus the filament load; **e)** Induced load per unit power consumption versus the filament load. Adjusted parameters used in simulations are listed in Table 2 except $\xi$.

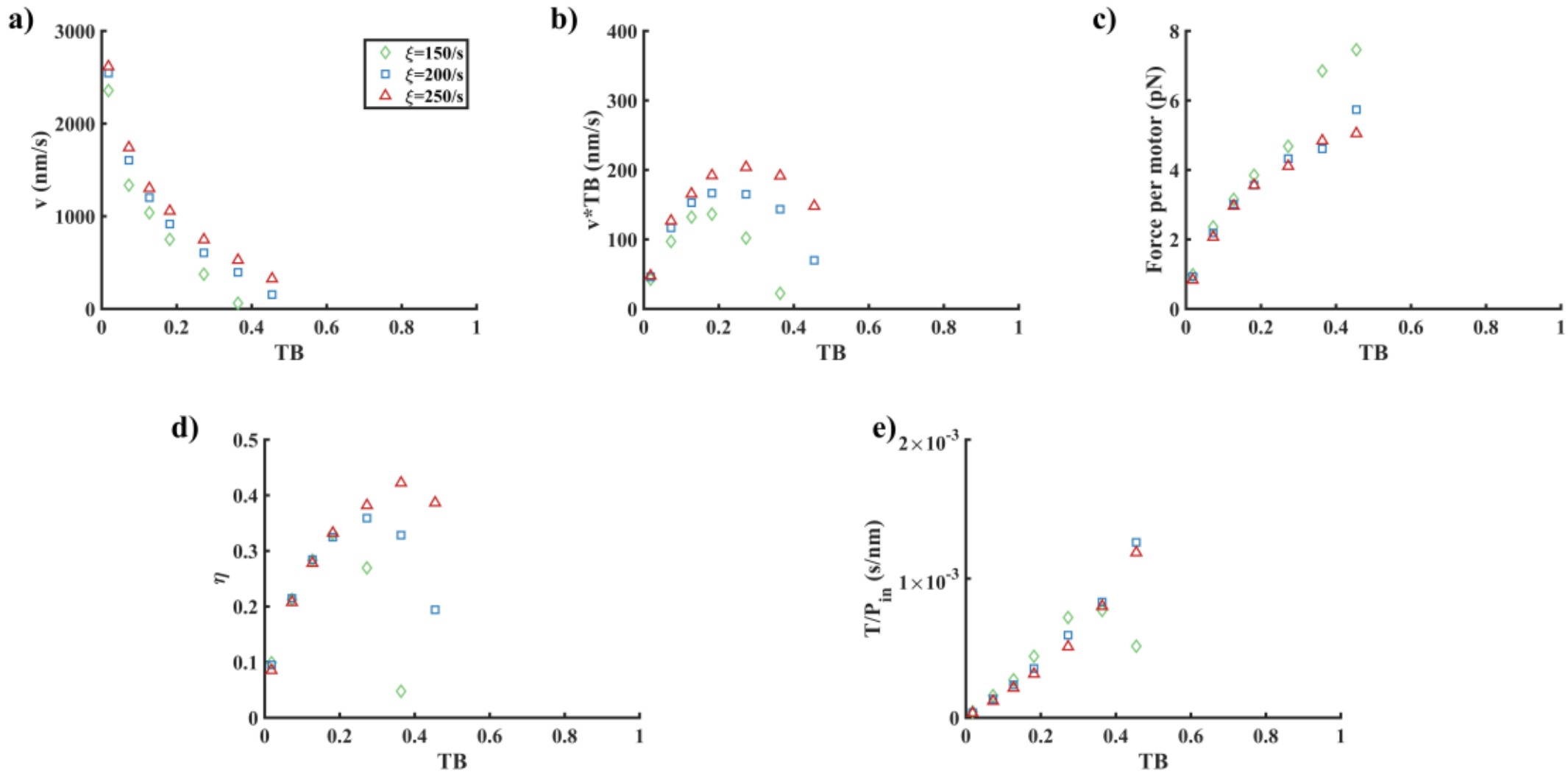

**Fig. S5** Predicted effects of $\xi$ on features of muscle contraction at constant [Pi] ([Pi] = 10mM): **a)** The shortening velocity versus the filament load; **b)** The output power versus the filament load; **c)** Force per working motor versus the filament load; **d)** The energy efficiency versus the filament load; **e)** Induced load per unit power consumption versus the filament load. Adjusted parameters used in simulations are listed in Table 2 except $\xi$.

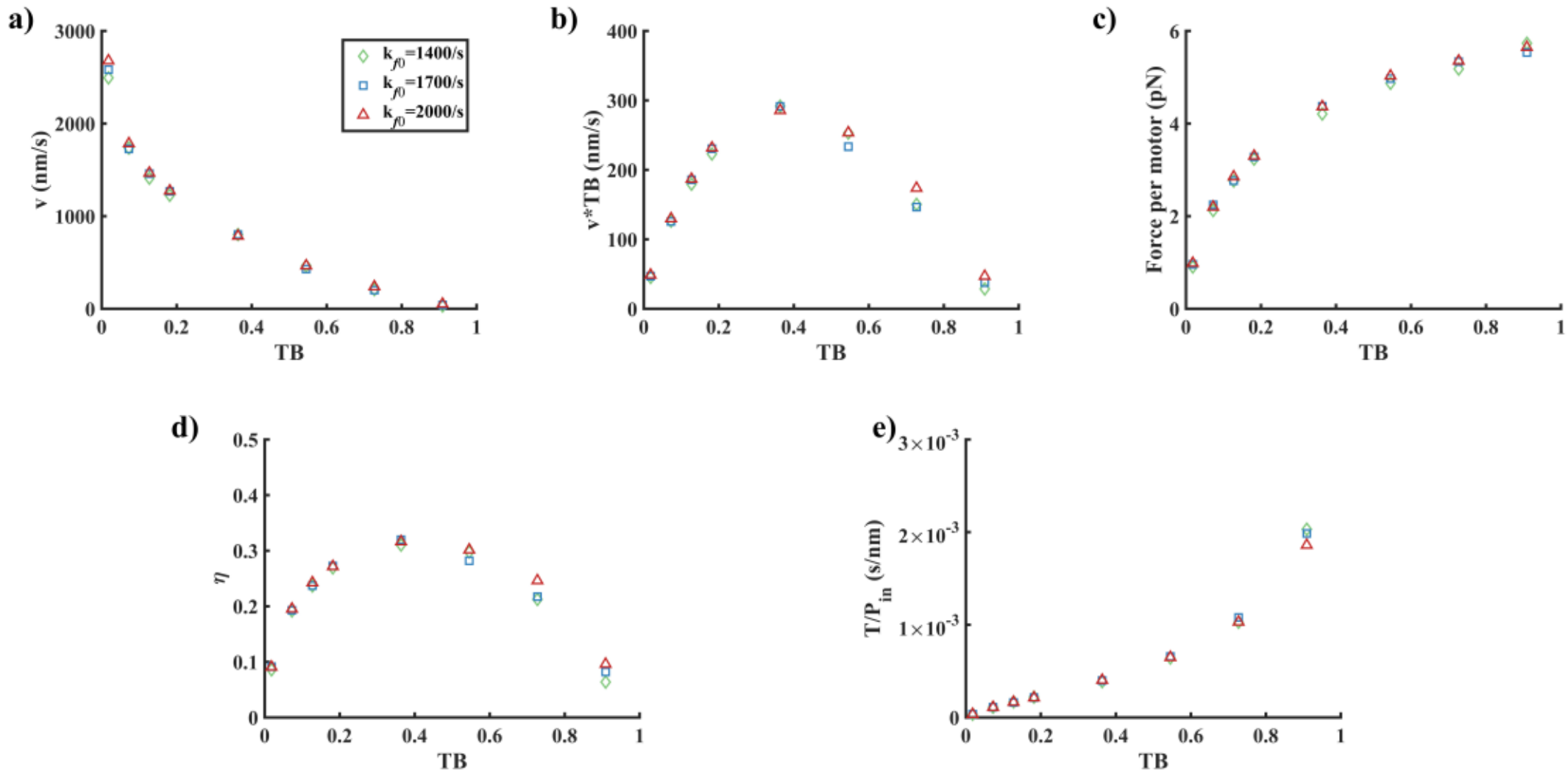

**Fig. S6** Predicted effects of $k_{f0}$ on features of muscle contraction at constant [Pi] ([Pi] = 1mM): **a)** The shortening velocity versus the filament load; **b)** The output power versus the filament load; **c)** Force per working motor versus the filament load; **d)** The energy efficiency versus the filament load; **e)** Induced load per unit power consumption versus the filament load. Adjusted parameters used in simulations are listed in Table 2 except $k_{f0}$.

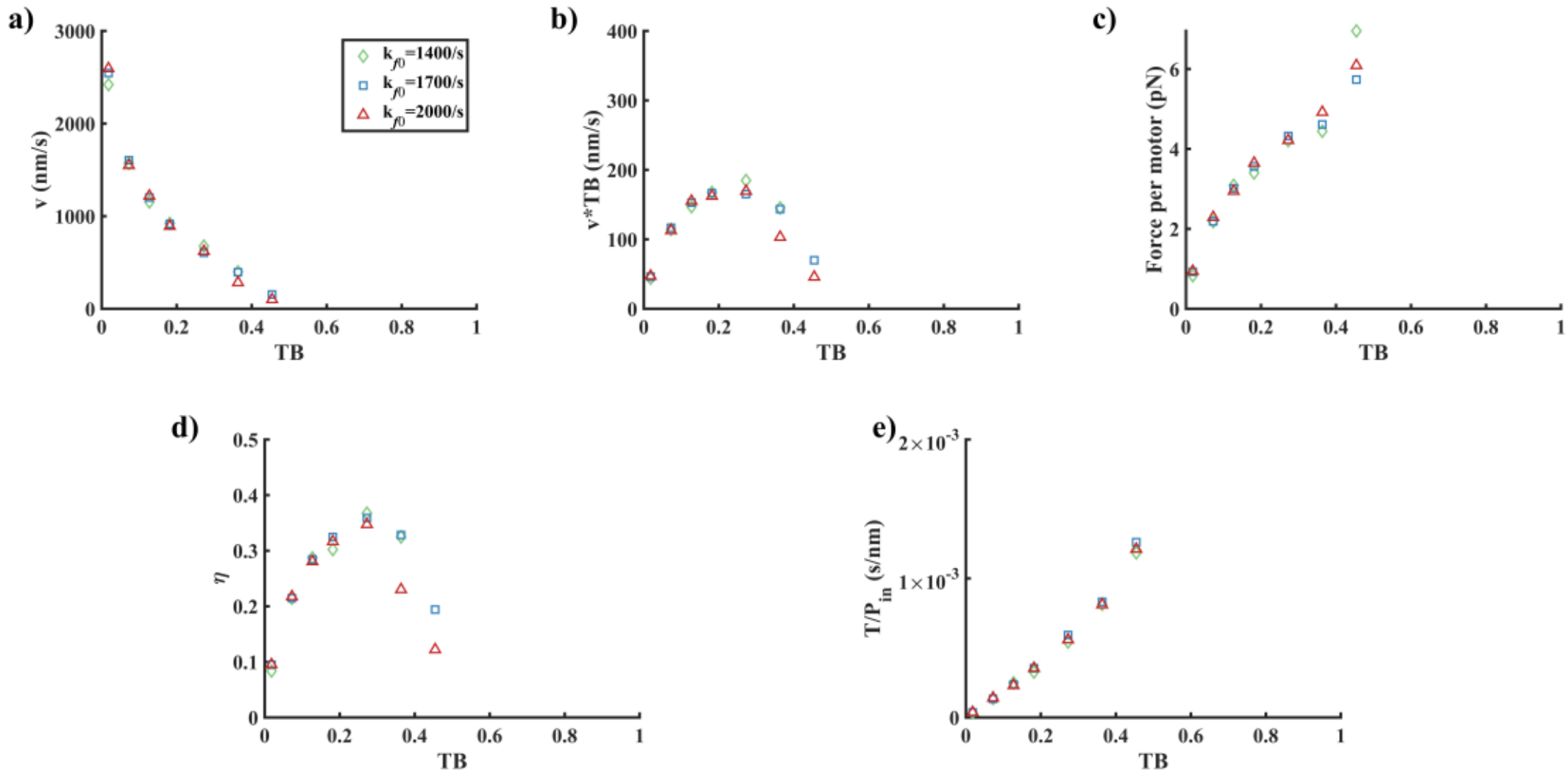

**Fig. S7** Predicted effects of $k_{f0}$ on features of muscle contraction at constant [Pi] ([Pi] = 10mM): **a)** The shortening velocity versus the filament load; **b)** The output power versus the filament load; **c)** Force per working motor versus the filament load; **d)** The energy efficiency versus the filament load; **e)** Induced load per unit power consumption versus the filament load. Adjusted parameters used in simulations are listed in Table 2 except $k_{f0}$.

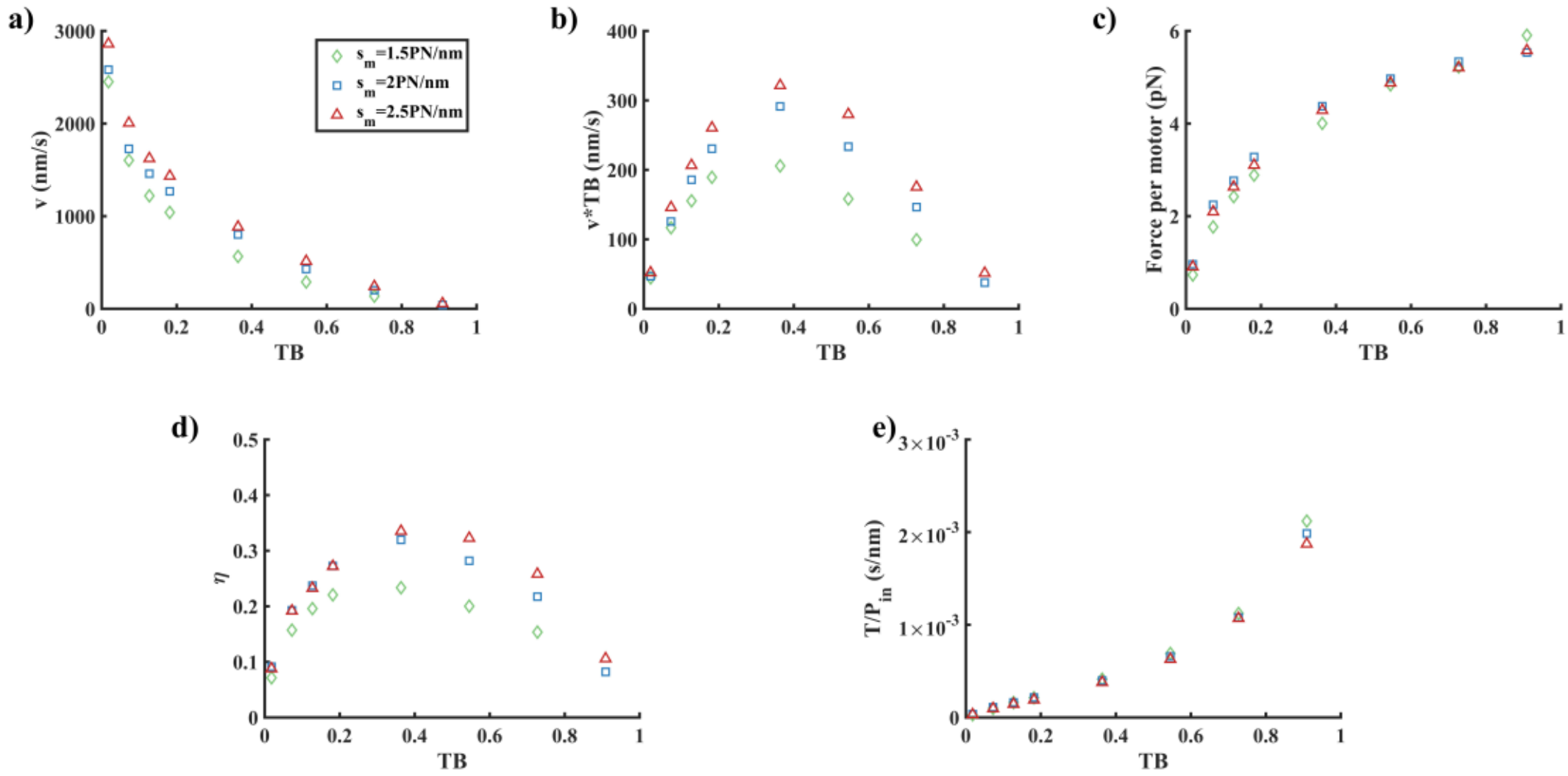

**Fig. S8** Predicted effects of $s_m$ on features of muscle contraction at constant [Pi] ([Pi] = 1mM): **a)** The shortening velocity versus the filament load; **b)** The output power versus the filament load; **c)** Force per working motor versus the filament load; **d)** The energy efficiency versus the filament load; **e)** Induced load per unit power consumption versus the filament load. Adjusted parameters used in simulations are listed in Table 2 except $s_m$.

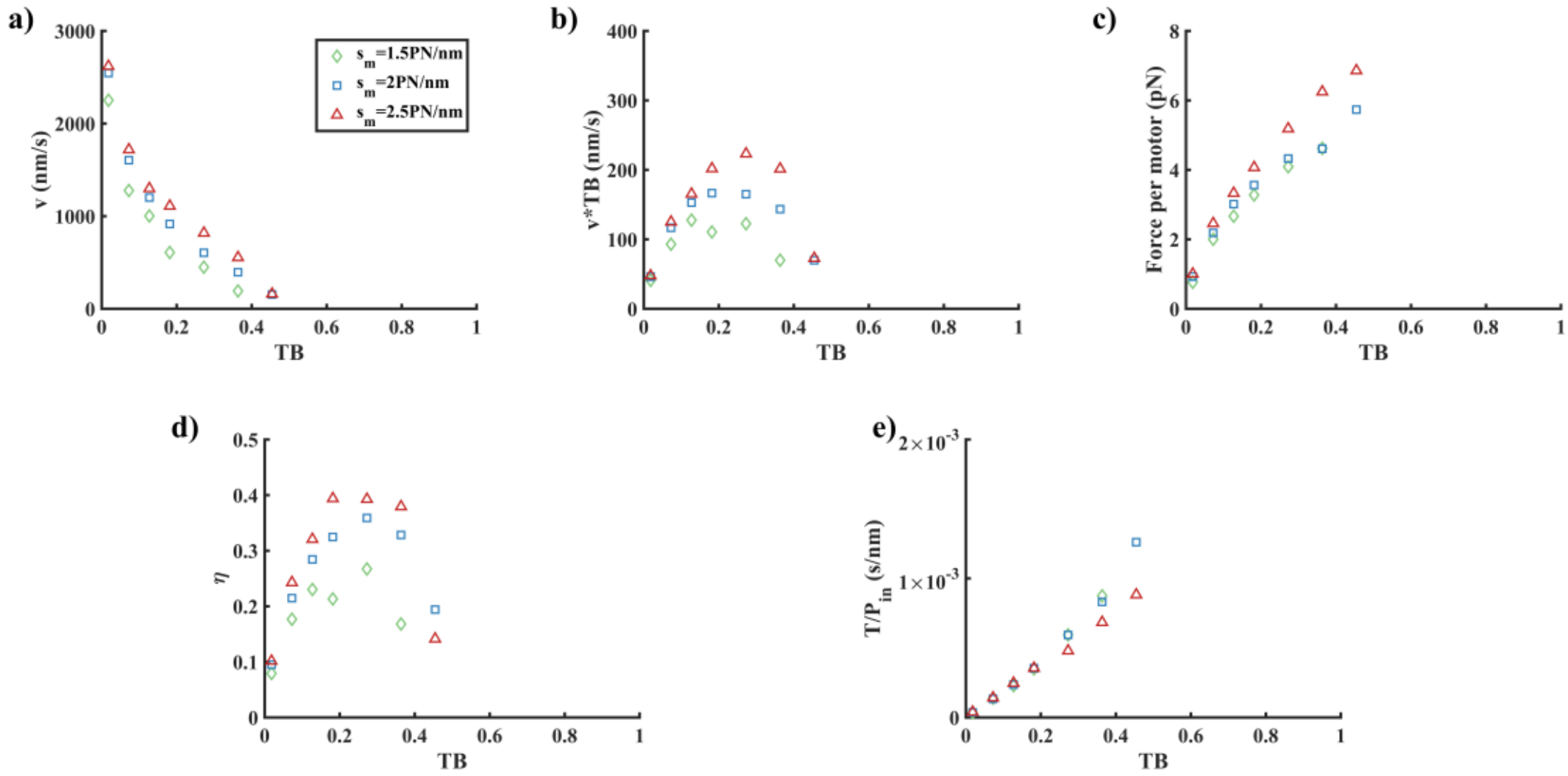

**Fig. S9** Predicted effects of $s_m$ on features of muscle contraction at constant [Pi] ([Pi] = 10mM): **a)** The shortening velocity versus the filament load; **b)** The output power versus the filament load; **c)** Force per working motor versus the filament load; **d)** The energy efficiency versus the filament load; **e)** Induced load per unit power consumption versus the filament load. Adjusted parameters used in simulations are listed in Table 2 except $s_m$.

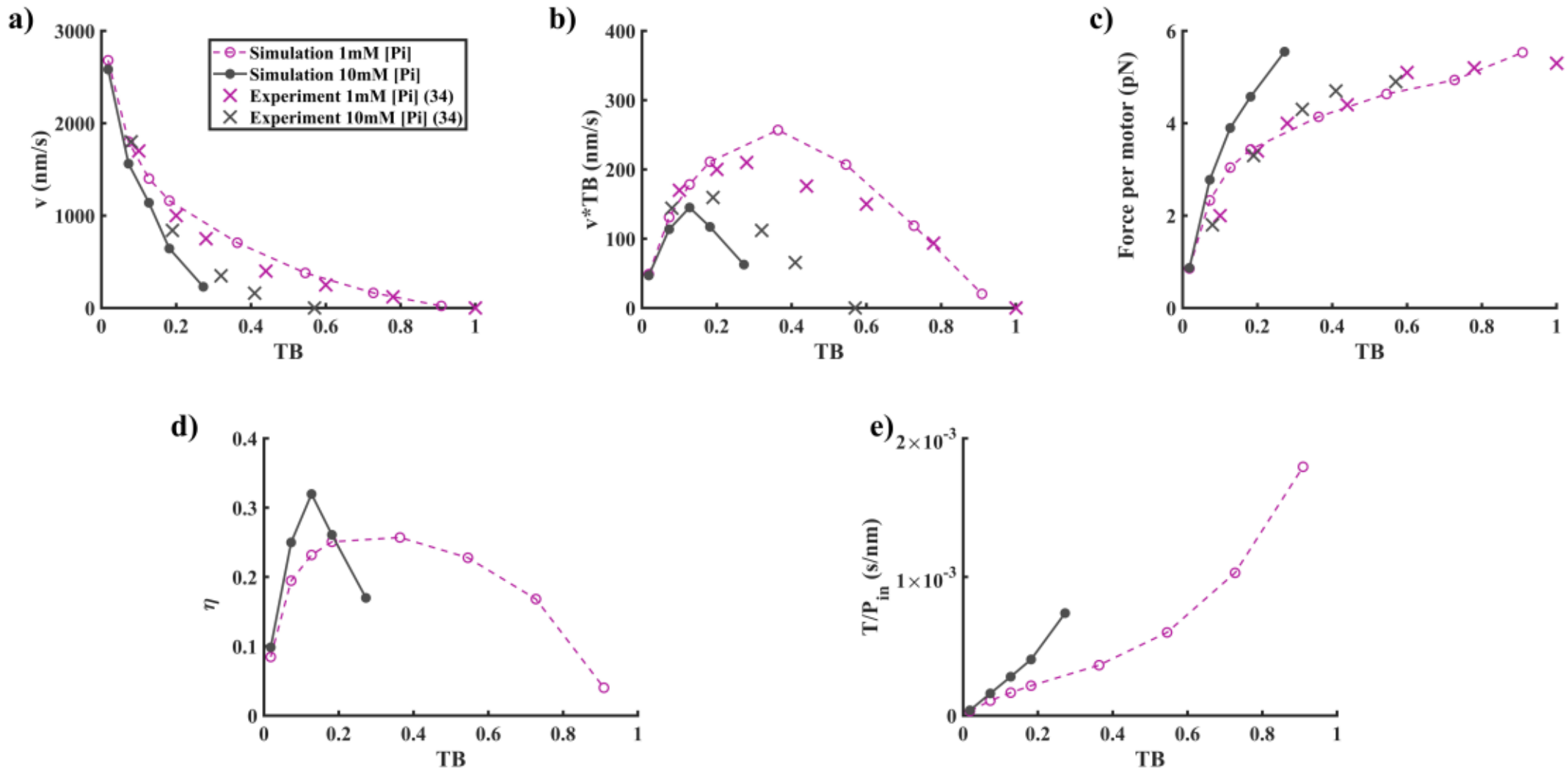

**Fig. S10** Predicted effects of a force-dependent Pi-release rate on features of muscle contraction at constant [Pi]: **a)** The shortening velocity versus the filament load; **b)** The output power versus the filament load; **c)** Force per working motor versus the filament load; **d)** The energy efficiency versus the filament load; **e)** Induced load per unit power consumption versus the filament load. In the simulations, the catch-slip bond breaking at AM*ADP.Pi is replaced with slip bond breaking and constant Pi-release rate is replaced with force dependent Pi-release, respectively. Adjusted bond breaking rates in the simulations are $k_{break1} = \begin{cases} 720/s & f \leq 0pN \\ 220/s \exp\left(\frac{f}{15}\right) & f > 0pN \end{cases}$, and $k_{break2} = 116.2/s \exp\left(-\frac{f}{2}\right) + 8.3/s \exp\left(\frac{f}{5}\right)$, respectively. Pi-release rate is given by $k_{Pi} = 250/s \exp\left(\frac{f}{5}\right)$. Other adjusted parameters used in simulations are listed in Table 2.

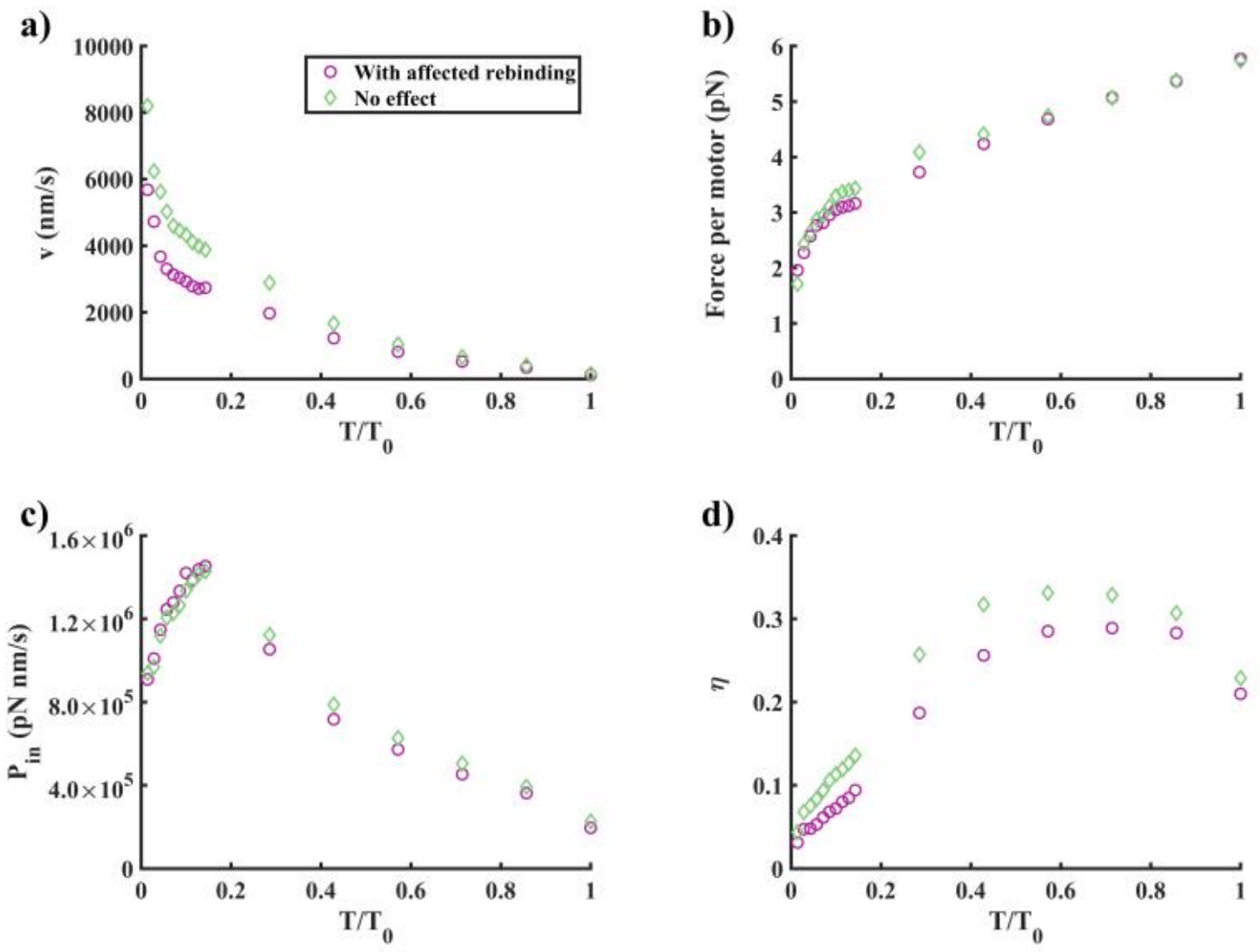

**Fig. S11** Predicted features of muscle contraction with or without affected rebinding rates of a detached Pi-bound myosin: a) Shortening velocity versus the filament load; b) Force per working motor versus the filament load; c) Power consumption versus the filament load; d) Energy efficiency versus the filament load. In the simulation, the rebinding rate is reduced 50% when the accumulated swing distance is above 3 nm.